\pdfoutput=1
\documentclass[aps,prl,amsmath,preprint,showpacs,superscriptaddress,
               letterpaper]{revtex4-1}
\usepackage{graphicx}
\usepackage{afterpage}
\usepackage{times}
\usepackage{siunitx}

\begin{document}
\preprint{Submitted to PRL}
\title{Molecular-scale remnants of the liquid-gas transition in supercritical polar fluids}

\author{V.~P.~Sokhan}
\affiliation{National Physical Laboratory, Hampton Road, Teddington, Middlesex TW11 0LW, UK}

\author{A.~Jones}
\affiliation{School of Physics and Astronomy, The University of Edinburgh,
             Mayfield Road, Edinburgh EH9 3JZ, UK}

\author{F.~S.~Cipcigan}
\affiliation{National Physical Laboratory, Hampton Road, Teddington, Middlesex TW11 0LW, UK}
\affiliation{School of Physics and Astronomy, The University of Edinburgh,
             Mayfield Road, Edinburgh EH9 3JZ, UK}

\author{J.~Crain}
\affiliation{National Physical Laboratory, Hampton Road, Teddington, 
             Middlesex TW11 0LW, UK}
\affiliation{School of Physics and Astronomy, The University of Edinburgh,
             Mayfield Road, Edinburgh EH9 3JZ, UK}

\author{G.~J.~Martyna}
\affiliation{IBM T.~J.~Watson Research Center, Yorktown Heights, New York 10598, USA}

\date{\today}
\begin{abstract}
An electronically coarse-grained model for water reveals a persistent vestige of the liquid-gas transition deep into the supercritical region. A crossover in the density dependence of the molecular dipole arises from the onset of non-percolating hydrogen bonds. The crossover points coincide with the Widom line in the scaling region but extend further, tracking the heat capacity maxima, offering evidence for liquid- and gas-like state points in a 
“one-phase” fluid. The effect is present even in dipole-limit models suggesting that it is common for all molecular liquids exhibiting dipole enhancement in the liquid phase.
\end{abstract}


\maketitle

There is a distinct boundary between liquid and vapor phases of a substance 
separated by a coexistence line over a finite range of pressures ($p$) and 
temperatures ($T$). Crossing the line results in a discontinuous density 
change -- the hallmark of a first-order phase transition. The coexistence 
line terminates at a critical point in the $(T,p)$ plane beyond which the 
thermodynamic distinction between liquid and gas phases is lost and a 
single-phase supercritical fluid is formed. In this region, there are no 
further phase boundaries and the supercritical fluid permits a continuous 
path from liquid to gas over a broad range of thermodynamic states \cite{Stan71}. 
In water, for example, this tunability can be exploited to control solvation 
properties making supercritical water an important ingredient in industrial 
processes and a promising ``green'' alternative to chemical solvents 
\cite{Arai02,Akiya02,Ecke96}.

While the classical thermodynamic picture is well understood, the molecular 
nature of supercritical fluids remains a largely open question. 
Key issues for water are (1) structure and hydrogen bonding (HB) in the 
supercritical region; (2) the electronic redistribution which occurs in the 
water molecules as the HB network is reconfigured; and (3) most broadly, the 
relationship of the supercritical fluid to fundamental liquid and gaseous 
states of matter that exist as separate phases below $T_c$.

In this context, the observation that certain thermophysical response 
functions exhibit maxima in the transition region between gas-like 
and liquid-like states of the supercritical phase 
has prompted the suggestion that a remnant of coexistence may extend into the 
supercritical region \cite{Braz11a}. 
The locus of such maxima is referred to as the Widom line \cite{Xu05}. 
Direct confirmation of distinct gas-like and liquid-like state points, 
however, is lacking though some evidence of dynamical crossover phenomena 
has been reported in noble gases \cite{Sime10,Braz11a}. 
Analogous concepts may apply at low temperatures where a Widom line is 
proposed to extend from a hidden liquid-liquid critical point \cite{Xu05} 
influencing the properties of supercooled water. 

Supercritical water poses unique challenges for molecular models because 
conventional empirical interaction potentials capture neither the electronic 
redistribution which occurs along an isotherm within the supercritical region 
of the phase diagram (dipole, quadrupole and higher manybody polarization responses)   
nor the manybody dispersion forces which have emerged as an important 
aspect of water physics \cite{Gill13,Sokh15}. Moreover, as these empirical 
models are typically optimized to describe liquid water near ambient 
conditions, their transferability to the supercritical fluid is 
questionable and the resulting physical insight into the molecular 
processes which occur above the critical point is limited. 

A new approach for materials simulation has been proposed recently wherein 
the molecular electronic distribution is not represented by a parametrized 
mean-field model, but rather by a coarse-grained description based on 
embedded harmonic oscillators -- the quantum Drude oscillator (QDO)
model \cite{Whit06,Whit07}. As the oscillators are treated quantum mechanically, 
this model generates the complete hierarchy of many-body polarisation and 
dispersion phenomena which when solved in strong coupling, as here, ensures that 
all potential symmetry-breaking interactions are present within Gaussian 
statistics.
The responses (to all orders) include, but are not limited to, inductive 
responses leading to distortions of the electronic charge distribution and 
to van der Waals and higher-order dispersion interactions arising from 
quantum mechanical charge density fluctuations. The model parameters are 
fit to leading-order responses of the isolated molecule providing a 
perfect low-density limit \cite{Sokh15,Jones13,Jones13a}. Here we apply the model 
with the short-range repulsion parameterized to a high-level ab initio 
calculations of the water dimer, (see Refs~\cite{Sokh15,SuppMat} for model details). 

In prior work \cite{Sokh15}, using the methodology described 
in \cite{Whit06,Whit07,Jones13,Jones13a} 
to simulate the model, we have determined the location of its critical point as
$\{T_\text{c}=649(2)$~K, $\rho_\text{c}=0.317(5)$~g/cm$^3\}$, which is in accord 
with experiment $\{T_\text{c}=647.096$~K, $\rho_\text{c}=0.322$~g/cm$^3\}$ \cite{Wagn02}. 
The dielectric properties along coexistence as well as the orthobaric densities 
of both the gas and liquid branches are well described \cite{Sokh15}.
We regard the prediction of an accurate critical point as a prerequisite for 
reliable examination of the supercritical phase. We are aware of no other 
description able to predict an accurate critical point from the properties 
of the isolated molecule.

Having established the critical point and the dielectric properties of the 
model, we are now in position to examine the molecular properties across 
several supercritical isotherms and to draw broader conclusions about the 
molecular nature of supercritical water. To set the context for this part 
of the work, we draw attention to recent studies which have explored the 
extent to which the notion of liquid-gas coexistence can be extended past 
the critical point. Typically, this extrapolation is based on identifying 
the loci of maxima in second-order thermodynamic response functions such 
as heat capacity ($C_p$), thermal expansivity ($\alpha$) or isothermal 
compressibility ($\kappa_T$) along isothermal paths. In atomic systems 
(noble gases), evidence of dynamic anomalies near the points of maxima in 
$C_p$ has been reported \cite{Sime10} and interpreted as a signature of 
liquid- and gas-like behavior. 
However, it is clear that the extrema of different thermodynamic functions 
rapidly diverge from each other above the critical point and only collapse 
onto a universal Widom line in the asymptotic scaling region near the 
critical point as described by a scaling theory developed 
in Ref.~\cite{Stan71}. 
Beyond this region, and further into the one-phase supercritical field, 
it is unclear whether any specific thermodynamic observable separates 
distinct regimes which can be unambiguously associated with 
liquid- or gas-like behavior.  

\begin{figure}[p]    %----------------------------------------- Figure 1 ------
  \includegraphics[width=86mm]{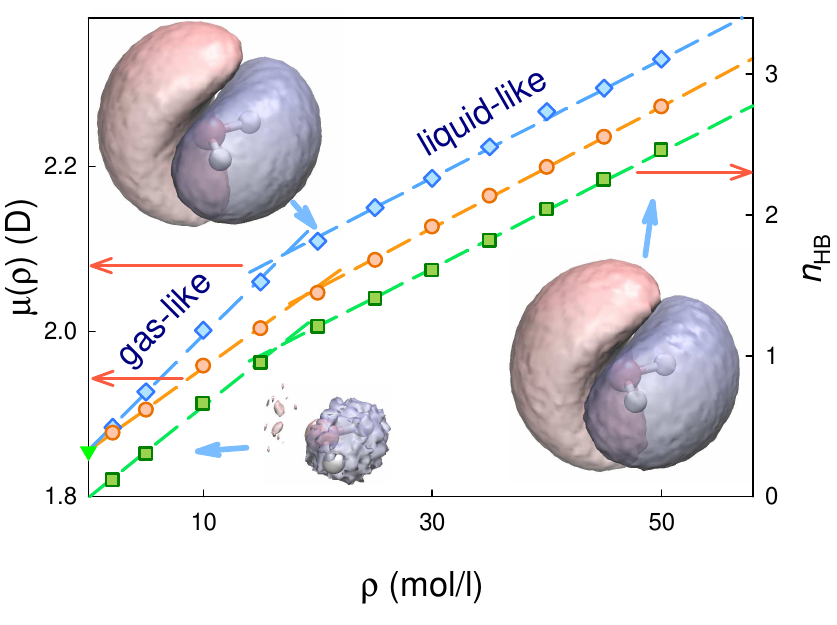}
  \caption{(Color online). Average molecular dipole moment as a function of 
	 density at $T=673$~K, blue diamonds, and at $T=873$~K, orange circles. The triangle 
	 denotes the isolated monomer value (parameter of the model). Green squares and 
	 right scale, average number of hydrogen bonds per molecule as a function 
	 of density at $T=673$~K. The lines are the linear fits to data.
	 Insets illustrate the variation in electron density with pressure, where 
	 the pink and blue isosurfaces correspond to gain and loss of the electron 
	 density.
  }
	\label{Figs:Dip}
\end{figure}

We now consider the dipole moment of water as a local reporter of molecular 
environment. Individual water molecules which are adaptive and responsive 
to their environment, become strongly polarized as a result of the highly 
directional bonding of the liquid-like state \cite{Bern33}. 
We explored the supercritical region along several isotherms from 673~K 
to 1083~K and for densities from 2~mol/l to 60~mol/l spanning 
from 10~MPa to 2.4~GPa. The results of our calculations of the molecular 
dipole at two isotherms in the supercritical region of the phase diagram are 
shown in Fig.~\ref{Figs:Dip}. The dipole's density dependence exhibits a 
change in slope at $\rho = 17.1$~mol/l at 673~K and 18.9~mol/l at 873~K. 
The corresponding densities of heat capacity maxima, obtained from the 
IAPWS-95 reference equation of state for water \cite{Wagn02}, are 
17.25~mol/l and 18.91~mol/l. The dipole crossover is still detectable at 
$T = 1083\,\text{K} = 1.7\times T_c$ above which it disappears in 
statistical noise. It becomes more pronounced with decreasing 
temperature and below $T_\text{c}$ it occurs in the two-phase region.

In order to investigate further the crossover in the dipole 
moment we analyze another microscopic discriminator between the 
liquid-like and gas-like states -- hydrogen bonding (HB).
Although the presence of a HB network in the supercritical region is 
well established by various techniques \cite{Hoff97,Wern05,Bern08a}, 
the structure and the topology of the network in liquid water is a matter 
of discussion \cite{Wern04,Head06,Bern08a,Sun14}. 
Examining HB connectivity, we find that local tetrahedral order persists 
down to low gas-like densities at supercritical temperatures. 
Using the distance-angle geometric definition of the hydrogen 
bond of Ref.~\cite{Kuma07} we observe a similar, albeit less pronounced, 
crossover in density dependence of the average number of hydrogen bonds per 
molecule, which occurs at the same densities as the crossover in dipole 
moment, shown in Fig.~\ref{Figs:Dip}.

The number of hydrogen bonds per molecule at 673~K and density $\rho=30$~mol/l 
(0.54 g/cm$^3$) is 40\% of that of ambient water, 300~K, 
55.32~mol/l, where $n_\text{HB}=3.71$. This can be compared with 
29\% of ambient at the same temperature and 28.9~mol/l estimated from  
NMR chemical shift measurements \cite{Hoff97}.
The number of hydrogen bonds per molecule at the crossover density, which 
decreases slightly from 1.08 at $T=673$~K to 0.85 at $T=1083$~K, is
well below the percolation threshold which occurs at 
$n_\text{HB}=1.53(5)$ according to Ref.~\cite{Blum84}.
This observation is in accord with Raman experiments \cite{Sun14}, which 
are interpreted to indicate no tetrahedral hydrogen bonding at the critical 
point. We note that collective properties which may show dynamic or viscoelastic 
anomalies may also be detectable by x-ray scattering techniques or other 
experimental probes of relaxation processes.  
We therefore conclude that the onset of the formation of minimal HB 
associations corresponds to the dipole crossover point in the 
supercritical field which, with decreasing temperatures, evolves to the 
liquid-gas coexistence line in the two phase region. 

\begin{figure}[p]    %----------------------------------------- Figure 2 ------
  \includegraphics[width=86mm]{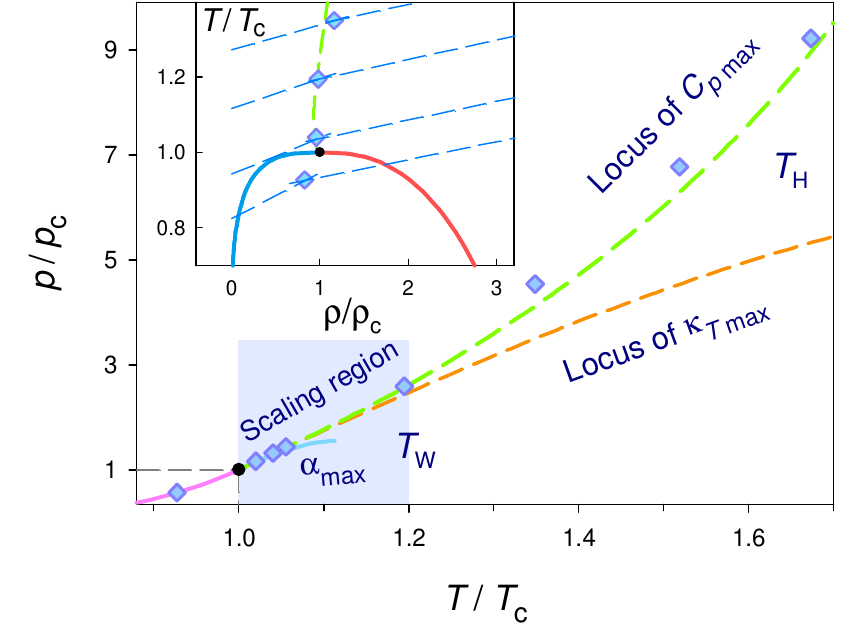}
  \caption{(Color online). $(T,p)$ phase diagram of water in the proximity of 
	the critical point in units of critical temperature, $T_\text{c}$, and critical 
	pressure, $p_\text{c}$. The liquid-vapor coexistence curve (pink), the loci of 
	maxima in heat capacity (green), thermal expansivity (orange), and isothermal 
	compressibility (blue), obtained using the IAPWS-95 reference EOS \cite{Wagn02}. 
	The blue diamonds represent our results derived from the analysis of Fig.~\ref{Figs:Dip}.
	The black dot marks the critical point. The inset, the corresponding $(\rho,T)$ 
	phase diagram in reduced units, shows the gas and liquid branches of the 
	coexistence curve, the locus of the heat capacity maxima, and superimposed are 
	linear fits to the dipole moment (the blue dashed lines) in gas- and liquid-like 
	regions with vertical offsets to map the corresponding temperatures at the crossover.  
	The crossover densities are denoted by the diamonds. The scaling theory of 
	Ref.~\cite{Stan71} predicts the ``true Widom line'' for water where 
	$\partial p/\partial T>0$ follows $\kappa_T$.}
	\label{Figs:Widom}
\end{figure}

The results of the study are summarized in Fig.~\ref{Figs:Widom}, 
where we plot the positions of the observed crossover points in the 
$(T,p)$ plane along with the portion of the coexistence line ending 
in the critical point and experimental data showing maxima in 
various response functions. Shown here are specific heat, $C_p$, 
isothermal compressibility, $\kappa_T$, and thermal expansivity, 
$\alpha$. Asymptotically close to the critical point, in the region 
$T_\text{c} < T < T_\text{W}$, the dipole crossover points occur 
along the Widom line, where the loci of all response function maxima 
coincide. Moreover, the dipole values are unambiguously vapor-like 
on the low-$p$ side and liquid-like on the high-$p$ side. On the 
molecular scale it therefore appears possible to partially extend 
the notion of coexistence and the liquid--gas transition well into 
the thermodynamically single-phase fluid as a remnant.

At higher temperatures, in the non-asymptotic region 
$T_\text{W} < T < T_\text{H}$, the loci of response function maxima diverge 
from the Widom line but it appears that the dipole crossover most closely 
tracks $C_p(T)$. Above $T_\text{H}$ the dipole derivative transition 
becomes undetectable and fluid appears homogeneous at all lenthscales.  
We note that the formation of HB dimers creates an additional thermal 
reservoir which contributes to the specific heat in liquid water. 
We therefore speculate that this may be the molecular origin of the 
close link revealed here between dipole-crossover, hydrogen bond formation, 
remnant liquid-gas coexistence behavior in the supercritical phase and the 
locus of maxima in $C_p$ in and beyond the scaling region. No such direct 
link exists between HB formation and thermal expansivity or isothermal 
compressibility the maxima of which follow very different trajectories 
out of the scaling region.

\begin{figure}[p]    %----------------------------------------- Figure 3 ------
  \includegraphics[width=86mm]{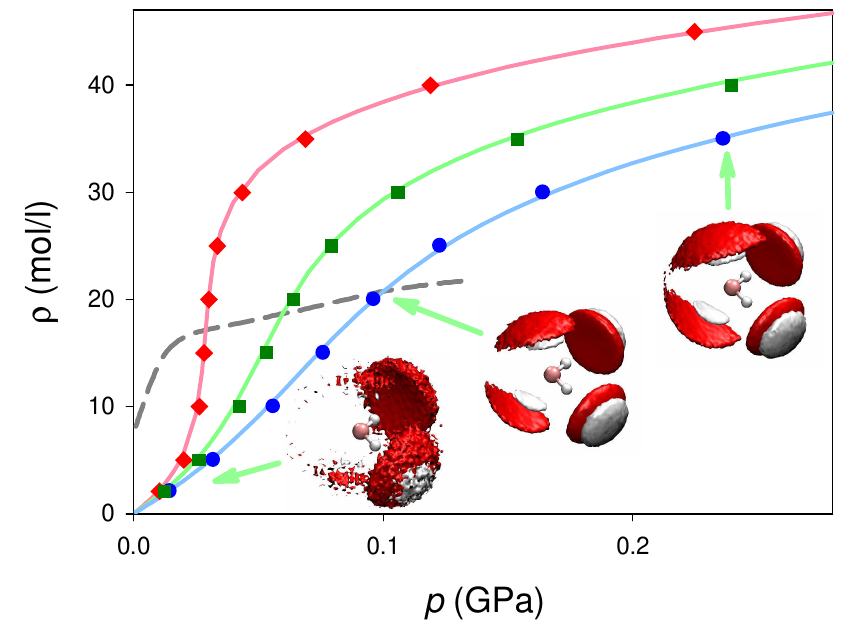}
  \caption{(Color online). Density of water along the $T=673$~K, 773~K, and 873~K 
	 isotherms (from top to bottom) as obtained in APIMD-QDO simulation	(symbols) vs 
	 experimental data(full lines). All simulated states, apart from the first four in 
	 each isotherm, are in the conventional supercritical region \cite{Arai02}. The 
	 dashed line indicates the crossover densities. Insets, the 3D plots showing first 
	 oxygen (red) and hydrogen (white) coordination shells, illustrate remarkable 
	 persistence of tetrahedral structure at $T=673$~K.}
	\label{Figs:Dens}
\end{figure}

Finally, we examine the structure and density at several thermodynamic state 
points for which experimental data are available \cite{Soper00}. 
We focus first on the $T=673$~K isotherm and compare the results to neutron 
diffraction measurements \cite{Soper00}. Predicted densities along this 
isotherm are shown in Fig.~\ref{Figs:Dens} together with data obtained from 
IAPWS-95 reference equation of state \cite{Wagn02}. Since our QDO model 
does not involve any condensed-phase parameterisation, the agreement arises 
as a prediction and gives confidence in the previously presented results.

\begin{figure}[p]    %----------------------------------------- Figure 4 ------
  \includegraphics[width=86mm]{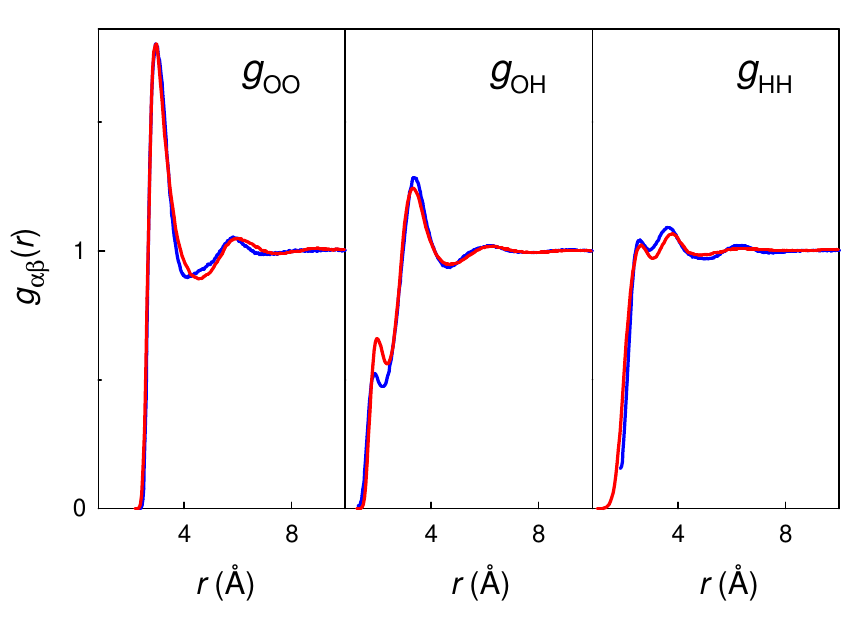}
  \caption{(Color online). The partial radial distribution functions obtained in the QDO water model 
	 simulation under supercritical conditions at $T=673$~K and $p=340$~MPa (red lines). 
	 Also shown for comparison are distribution functions obtained from neutron 
	 scattering \cite{Soper00} at the same thermodynamic conditions (blue lines).}
	\label{Figs:RDF}
\end{figure}

Further structural data is presented in the insets to Fig.~\ref{Figs:Dens},
which show three-dimensional first coordination shell plots for oxygen 
(red) and hydrogen (white) of surrounding water molecules. The isosurfaces 
are drawn at $2\times$ (bulk density) levels.
These illustrate the extent of hydrogen bond reconfiguration 
which occurs at various densities along the isotherm.  
The partial radial distribution functions are shown in Fig.~\ref{Figs:RDF}
for $T=673$~K, $p=340$~MPa (red lines) in comparison to the neutron 
diffraction data of Ref.~\cite{Soper00} (blue lines). 
Subtle features of the experimental H--H distributions are 
present in the model, and the O--O correlations are captured as well.

In Fig.~\ref{Figs:Dip}, the mean molecular dipole moment $|\bar{\mu}|$ 
is seen to vary linearly with density in two regions, which can be 
understood using the following argument: The induced dipole moment is 
proportional to the local field acting on a water molecule, which can 
be decomposed into contributions from the first coordination shell and 
from the rest of the fluid. As ab initio and induction model calculations 
of small water clusters \cite{Greg97} have shown, the water dipole moment 
scales linearly with the cluster size for up to five members in the 
cluster. A mean field estimate assuming the average distance to scale 
like $\rho^{-1/3}$ leads to the same conclusion. More deeply, the 
Onsager reaction field strength scales linearly with density \cite{Onsa36}.

For strongly associated liquids like water, the Widom line has been connected 
with a percolation transition in the hydrogen-bond network \cite{Part07}. 
However, our results support the view that the origin of the crossover and 
the extrema in the response functions is more local. We show that this 
behavior extends much further into the notionally one-phase region and can 
be detected as far as $T = 1.7\times T_\text{c}$ tracking most closely 
the maxima in the specific heat. Only along the isotherms beyond these 
temperatures does it appear that supercritical water is a homogeneous fluid on 
the molecular scale exhibiting no detectable remnants of the liquid-gas transition. 
The extent of hydrogen bonding across the Widom line and beyond the scaling 
region can, in principle, be indirectly accessed by Raman scattering 
measurements of the OH stretch band which is a reporter of hydrogen
bonding and the integrity of the network. Collective properties which may 
show dynamic or viscoelastic anomalies may also be detectable by x-ray 
scattering techniques or other experimental probes of relaxation processes.

\begin{figure}[p]    %----------------------------------------- Figure 5 ------
  \includegraphics[width=86mm]{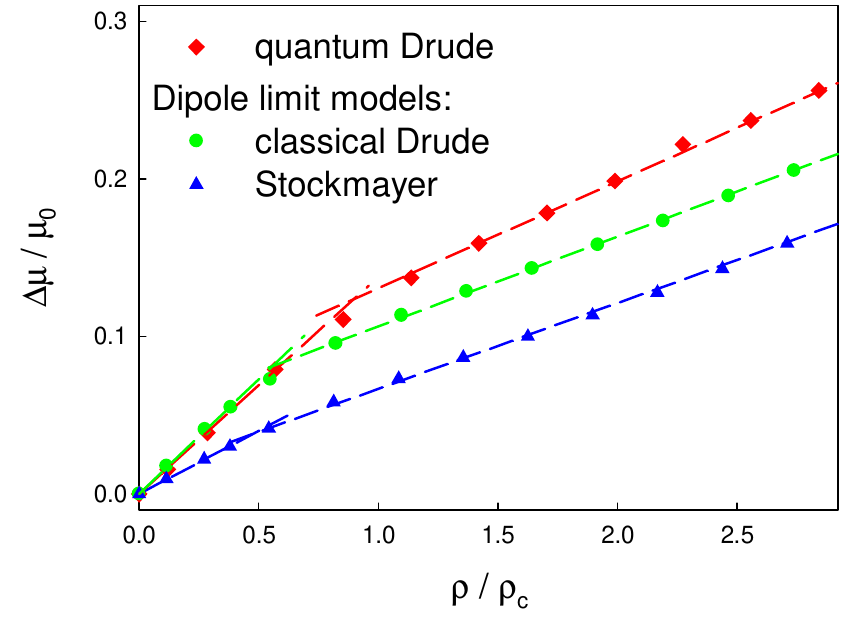}
  \caption{(Color online). Relative (to the gas-phase value) enhancement of the molecular dipole  
	 moment as a function of the reduced density. Red diamonds, quantum Drude model; 
	 green circles, classical Drude oscillator model (dipole limit); 
	 blue triangles, polarizable Stockmayer model (dipole limit).
	 All calculations were performed at a reduced temperature $T^*=1.036\,T_\text{c}$ in terms of 
	 corresponding critical temperature $T_\text{c}$. Straight lines are to guide the
	 eye.}
	\label{Figs:Dipoles}
\end{figure}

In order to address the question of generality of the phenomena reported 
here we studied two dipole polarizable models (polarization treated 
in the electric dipole limit), namely the classical Drude oscillator 
model, SWM4-NDP \cite{Lamo06}, for which the critical parameters are 
known \cite{Kiss12}: $T^\text{dip}_\text{c} = 0.89T_\text{c}$, 
$p^\text{dip}_\text{c} = 0.90p_\text{c}$, 
$\rho^\text{dip}_\text{c} = 1.02\rho_\text{c}$, in terms of experimental 
values for water. The second is a polarizable version of the Stockmayer 
model, the `minimalist' (purely dipolar) fluid model with a different 
local association topology \cite{Leeu94}, and estimated critical parameters 
$T^\text{s}_\text{c} = 1.1T_\text{c}$, $p^\text{s}_\text{c} = 1.5p_\text{c}$, 
$\rho^\text{s}_\text{c} = 1.02\rho_\text{c}$ (see Ref.~\cite{SuppMat}). 
Finite size effects are investigated and ruled out in Supplementary 
Material~\cite{SuppMat}. The results for three models, presented in 
Fig.~\ref{Figs:Dipoles}, show that the dipole moment change is most 
pronounced in fully responsive QDO, but also emerges from dipole-polarizable 
water model and the polarizable Stockmayer model. The reduced crossover 
density for QDO model $\rho^*_\times=0.90$, and shifts to 0.63 for SWM4-NDP 
and to 0.56 for the Stockmayer model. Therefore, our study has revealed an 
intriguing new property of associating liquids, the extension of gas-liquid 
critical effects on the molecular dipole moment.
%, following the maximum in the heat capacity out of the scaling regime.

In summary, we find direct evidence of molecular-scale heterogeneity in 
supercritical polar fluids. A clearly identifiable transition between dissociated 
(gas like) and associated (liquid like) regimes is evidenced by a change in the 
density dependence of the molecular dipole moment accompanied by a transition in 
the hydrogen bond connectivity.   
This characteristic signature has been observed in a fully responsive electronic 
model of water which provides an excellent prediction of the ambient, critical 
and supercritical properties but is also present in simpler dipole-limit models 
with two different association topologies -- the minimal model class in which 
the the behavior may be observed. While the simpler models do not predict the 
critical point accurately for water they nevertheless provide compelling 
evidence that the phenomenon is likely to be general for polar liquids. 
The observations provide a molecular basis for the variety of response function 
anomalies which define the ``Widom line'' asymptotically close to the critical 
point. This is also significant outside the scaling regime where the maxima of 
thermodynamic response functions follow different trajectories making it unclear 
which, if any, separate liquid-like from gas-like regions. Here we observe 
that the dipole crossover points follow only the heat capacity maximum outside 
of the scaling region and their correlation to rudimentary hydrogen bonding 
establishes an unambiguous extension of the liquid-gas coexistence line 
deep into the supercritical phase.

This work was supported by the NPL Strategic Research programme, the Engineering 
and Physical Sciences Research Council (EPSRC) and the European Metrology Research 
Programme (EMRP). Generous allocation of time on BG/Q at STFC Hartee Centre, UK, is 
gratefully acknowledged.
%--------------------------------------------------------------------------------
\bibliography{LY14576}

\begin{thebibliography}{31}%
\makeatletter
\providecommand \@ifxundefined [1]{%
 \@ifx{#1\undefined}
}%
\providecommand \@ifnum [1]{%
 \ifnum #1\expandafter \@firstoftwo
 \else \expandafter \@secondoftwo
 \fi
}%
\providecommand \@ifx [1]{%
 \ifx #1\expandafter \@firstoftwo
 \else \expandafter \@secondoftwo
 \fi
}%
\providecommand \natexlab [1]{#1}%
\providecommand \enquote  [1]{``#1''}%
\providecommand \bibnamefont  [1]{#1}%
\providecommand \bibfnamefont [1]{#1}%
\providecommand \citenamefont [1]{#1}%
\providecommand \href@noop [0]{\@secondoftwo}%
\providecommand \href [0]{\begingroup \@sanitize@url \@href}%
\providecommand \@href[1]{\@@startlink{#1}\@@href}%
\providecommand \@@href[1]{\endgroup#1\@@endlink}%
\providecommand \@sanitize@url [0]{\catcode `\\12\catcode `\$12\catcode
  `\&12\catcode `\#12\catcode `\^12\catcode `\_12\catcode `\%12\relax}%
\providecommand \@@startlink[1]{}%
\providecommand \@@endlink[0]{}%
\providecommand \url  [0]{\begingroup\@sanitize@url \@url }%
\providecommand \@url [1]{\endgroup\@href {#1}{\urlprefix }}%
\providecommand \urlprefix  [0]{URL }%
\providecommand \Eprint [0]{\href }%
\providecommand \doibase [0]{http://dx.doi.org/}%
\providecommand \selectlanguage [0]{\@gobble}%
\providecommand \bibinfo  [0]{\@secondoftwo}%
\providecommand \bibfield  [0]{\@secondoftwo}%
\providecommand \translation [1]{[#1]}%
\providecommand \BibitemOpen [0]{}%
\providecommand \bibitemStop [0]{}%
\providecommand \bibitemNoStop [0]{.\EOS\space}%
\providecommand \EOS [0]{\spacefactor3000\relax}%
\providecommand \BibitemShut  [1]{\csname bibitem#1\endcsname}%
\let\auto@bib@innerbib\@empty
%</preamble>
\bibitem [{\citenamefont {Stanley}(1971)}]{Stan71}%
  \BibitemOpen
  \bibfield  {author} {\bibinfo {author} {\bibfnamefont {H.~E.}\ \bibnamefont
  {Stanley}},\ }\href@noop {} {\emph {\bibinfo {title} {Introduction to Phase
  Transitions and Critical Phenomena}}}\ (\bibinfo  {publisher} {Oxford
  University Press},\ \bibinfo {address} {New York},\ \bibinfo {year}
  {1971})\BibitemShut {NoStop}%
\bibitem [{\citenamefont {Arai}\ \emph {et~al.}(2002)\citenamefont {Arai},
  \citenamefont {Sako},\ and\ \citenamefont {Takebayashi}}]{Arai02}%
  \BibitemOpen
  \bibinfo {editor} {\bibfnamefont {Y.}~\bibnamefont {Arai}}, \bibinfo {editor}
  {\bibfnamefont {T.}~\bibnamefont {Sako}}, \ and\ \bibinfo {editor}
  {\bibfnamefont {Y.}~\bibnamefont {Takebayashi}},\ eds.,\ \href {\doibase
  10.1007/978-3-642-56238-9} {\emph {\bibinfo {title} {Supercritical Fluids}}}\
  (\bibinfo  {publisher} {Springer, Berlin--Heidelberg},\ \bibinfo {year}
  {2002})\BibitemShut {NoStop}%
\bibitem [{\citenamefont {Akiya}\ and\ \citenamefont {Savage}(2002)}]{Akiya02}%
  \BibitemOpen
  \bibfield  {author} {\bibinfo {author} {\bibfnamefont {N.}~\bibnamefont
  {Akiya}}\ and\ \bibinfo {author} {\bibfnamefont {P.~E.}\ \bibnamefont
  {Savage}},\ }\href {\doibase 10.1021/cr000668w} {\bibfield  {journal}
  {\bibinfo  {journal} {Chem.\ Rev.}\ }\textbf {\bibinfo {volume} {102}},\
  \bibinfo {pages} {2725} (\bibinfo {year} {2002})}\BibitemShut {NoStop}%
\bibitem [{\citenamefont {Eckert}\ \emph {et~al.}(1996)\citenamefont {Eckert},
  \citenamefont {Knutson},\ and\ \citenamefont {Debenedetti}}]{Ecke96}%
  \BibitemOpen
  \bibfield  {author} {\bibinfo {author} {\bibfnamefont {C.~A.}\ \bibnamefont
  {Eckert}}, \bibinfo {author} {\bibfnamefont {B.~L.}\ \bibnamefont {Knutson}},
  \ and\ \bibinfo {author} {\bibfnamefont {P.~G.}\ \bibnamefont
  {Debenedetti}},\ }\href {\doibase 10.1038/383313a0} {\bibfield  {journal}
  {\bibinfo  {journal} {Nature}\ }\textbf {\bibinfo {volume} {383}},\ \bibinfo
  {pages} {313} (\bibinfo {year} {1996})}\BibitemShut {NoStop}%
\bibitem [{\citenamefont {Brazhkin}\ \emph {et~al.}(2011)\citenamefont
  {Brazhkin}, \citenamefont {Fomin}, \citenamefont {Lyapin}, \citenamefont
  {Ryzhov},\ and\ \citenamefont {Tsiok}}]{Braz11a}%
  \BibitemOpen
  \bibfield  {author} {\bibinfo {author} {\bibfnamefont {V.~V.}\ \bibnamefont
  {Brazhkin}}, \bibinfo {author} {\bibfnamefont {Y.~D.}\ \bibnamefont {Fomin}},
  \bibinfo {author} {\bibfnamefont {A.~G.}\ \bibnamefont {Lyapin}}, \bibinfo
  {author} {\bibfnamefont {V.~N.}\ \bibnamefont {Ryzhov}}, \ and\ \bibinfo
  {author} {\bibfnamefont {E.~N.}\ \bibnamefont {Tsiok}},\ }\href {\doibase
  10.1021/jp2039898} {\bibfield  {journal} {\bibinfo  {journal} {J.\ Phys.\
  Chem.\ B}\ }\textbf {\bibinfo {volume} {115}},\ \bibinfo {pages} {14112}
  (\bibinfo {year} {2011})}\BibitemShut {NoStop}%
\bibitem [{\citenamefont {Xu}\ \emph {et~al.}(2005)\citenamefont {Xu},
  \citenamefont {Kumar}, \citenamefont {Buldyrev}, \citenamefont {Chen},
  \citenamefont {Poole}, \citenamefont {Sciortino},\ and\ \citenamefont
  {Stanley}}]{Xu05}%
  \BibitemOpen
  \bibfield  {author} {\bibinfo {author} {\bibfnamefont {L.}~\bibnamefont
  {Xu}}, \bibinfo {author} {\bibfnamefont {P.}~\bibnamefont {Kumar}}, \bibinfo
  {author} {\bibfnamefont {S.~V.}\ \bibnamefont {Buldyrev}}, \bibinfo {author}
  {\bibfnamefont {S.-H.}\ \bibnamefont {Chen}}, \bibinfo {author}
  {\bibfnamefont {P.~H.}\ \bibnamefont {Poole}}, \bibinfo {author}
  {\bibfnamefont {F.}~\bibnamefont {Sciortino}}, \ and\ \bibinfo {author}
  {\bibfnamefont {H.~E.}\ \bibnamefont {Stanley}},\ }\href {\doibase
  10.1073/pnas.0507870102} {\bibfield  {journal} {\bibinfo  {journal} {Proc.\
  Natl Acad.\ Sci.\ USA}\ }\textbf {\bibinfo {volume} {102}},\ \bibinfo {pages}
  {16558} (\bibinfo {year} {2005})}\BibitemShut {NoStop}%
\bibitem [{\citenamefont {Simeoni}\ \emph {et~al.}(2010)\citenamefont
  {Simeoni}, \citenamefont {Bryk}, \citenamefont {Gorelli}, \citenamefont
  {Krisch}, \citenamefont {Ruocco}, \citenamefont {Santoro},\ and\
  \citenamefont {Scopigno}}]{Sime10}%
  \BibitemOpen
  \bibfield  {author} {\bibinfo {author} {\bibfnamefont {G.~G.}\ \bibnamefont
  {Simeoni}}, \bibinfo {author} {\bibfnamefont {T.}~\bibnamefont {Bryk}},
  \bibinfo {author} {\bibfnamefont {F.~A.}\ \bibnamefont {Gorelli}}, \bibinfo
  {author} {\bibfnamefont {M.}~\bibnamefont {Krisch}}, \bibinfo {author}
  {\bibfnamefont {G.}~\bibnamefont {Ruocco}}, \bibinfo {author} {\bibfnamefont
  {M.}~\bibnamefont {Santoro}}, \ and\ \bibinfo {author} {\bibfnamefont
  {T.}~\bibnamefont {Scopigno}},\ }\href {\doibase 10.1038/nphys1683}
  {\bibfield  {journal} {\bibinfo  {journal} {Nat.\ Phys.}\ }\textbf {\bibinfo
  {volume} {6}},\ \bibinfo {pages} {503} (\bibinfo {year} {2010})}\BibitemShut
  {NoStop}%
\bibitem [{\citenamefont {Gillan}\ \emph {et~al.}(2013)\citenamefont {Gillan},
  \citenamefont {Alf{\`e}}, \citenamefont {Bart{\'o}k},\ and\ \citenamefont
  {Cs{\'a}nyi}}]{Gill13}%
  \BibitemOpen
  \bibfield  {author} {\bibinfo {author} {\bibfnamefont {M.~J.}\ \bibnamefont
  {Gillan}}, \bibinfo {author} {\bibfnamefont {D.}~\bibnamefont {Alf{\`e}}},
  \bibinfo {author} {\bibfnamefont {A.~P.}\ \bibnamefont {Bart{\'o}k}}, \ and\
  \bibinfo {author} {\bibfnamefont {G.}~\bibnamefont {Cs{\'a}nyi}},\ }\href
  {\doibase http://dx.doi.org/10.1063/1.4852182} {\bibfield  {journal}
  {\bibinfo  {journal} {J.\ Chem.\ Phys.}\ }\textbf {\bibinfo {volume} {139}},\
  \bibinfo {pages} {244504} (\bibinfo {year} {2013})}\BibitemShut {NoStop}%
\bibitem [{\citenamefont {Sokhan}\ \emph {et~al.}(2015)\citenamefont {Sokhan},
  \citenamefont {Jones}, \citenamefont {Crain}, \citenamefont {Cipcigan},\ and\
  \citenamefont {Martyna}}]{Sokh15}%
  \BibitemOpen
  \bibfield  {author} {\bibinfo {author} {\bibfnamefont {V.~P.}\ \bibnamefont
  {Sokhan}}, \bibinfo {author} {\bibfnamefont {A.}~\bibnamefont {Jones}},
  \bibinfo {author} {\bibfnamefont {J.}~\bibnamefont {Crain}}, \bibinfo
  {author} {\bibfnamefont {F.}~\bibnamefont {Cipcigan}}, \ and\ \bibinfo
  {author} {\bibfnamefont {G.}~\bibnamefont {Martyna}},\ }\href {\doibase
  10.1073/pnas.1418982112} {\bibfield  {journal} {\bibinfo  {journal} {Proc.\
  Natl Acad.\ Sci.\ USA}\ }\textbf {\bibinfo {volume} {112}},\ \bibinfo {pages}
  {6341} (\bibinfo {year} {2015})}\BibitemShut {NoStop}%
\bibitem [{\citenamefont {Whitfield}\ and\ \citenamefont
  {Martyna}(2006)}]{Whit06}%
  \BibitemOpen
  \bibfield  {author} {\bibinfo {author} {\bibfnamefont {T.~W.}\ \bibnamefont
  {Whitfield}}\ and\ \bibinfo {author} {\bibfnamefont {G.~J.}\ \bibnamefont
  {Martyna}},\ }\href {\doibase 10.1016/j.cplett.2006.04.035} {\bibfield
  {journal} {\bibinfo  {journal} {Chem.\ Phys.\ Lett.}\ }\textbf {\bibinfo
  {volume} {424}},\ \bibinfo {pages} {409} (\bibinfo {year}
  {2006})}\BibitemShut {NoStop}%
\bibitem [{\citenamefont {Whitfield}\ and\ \citenamefont
  {Martyna}(2007)}]{Whit07}%
  \BibitemOpen
  \bibfield  {author} {\bibinfo {author} {\bibfnamefont {T.~W.}\ \bibnamefont
  {Whitfield}}\ and\ \bibinfo {author} {\bibfnamefont {G.~J.}\ \bibnamefont
  {Martyna}},\ }\href {\doibase 10.1063/1.2424708} {\bibfield  {journal}
  {\bibinfo  {journal} {J.\ Chem.\ Phys.}\ }\textbf {\bibinfo {volume} {126}},\
  \bibinfo {eid} {074104} (\bibinfo {year} {2007})}\BibitemShut {NoStop}%
\bibitem [{\citenamefont {Jones}\ \emph
  {et~al.}(2013{\natexlab{a}})\citenamefont {Jones}, \citenamefont {Crain},
  \citenamefont {Sokhan}, \citenamefont {Whitfield},\ and\ \citenamefont
  {Martyna}}]{Jones13}%
  \BibitemOpen
  \bibfield  {author} {\bibinfo {author} {\bibfnamefont {A.~P.}\ \bibnamefont
  {Jones}}, \bibinfo {author} {\bibfnamefont {J.}~\bibnamefont {Crain}},
  \bibinfo {author} {\bibfnamefont {V.~P.}\ \bibnamefont {Sokhan}}, \bibinfo
  {author} {\bibfnamefont {T.~W.}\ \bibnamefont {Whitfield}}, \ and\ \bibinfo
  {author} {\bibfnamefont {G.~J.}\ \bibnamefont {Martyna}},\ }\href {\doibase
  10.1103/PhysRevB.87.144103} {\bibfield  {journal} {\bibinfo  {journal}
  {Phys.\ Rev.\ B}\ }\textbf {\bibinfo {volume} {87}},\ \bibinfo {pages}
  {144103} (\bibinfo {year} {2013}{\natexlab{a}})}\BibitemShut {NoStop}%
\bibitem [{\citenamefont {Jones}\ \emph
  {et~al.}(2013{\natexlab{b}})\citenamefont {Jones}, \citenamefont {Cipcigan},
  \citenamefont {Sokhan}, \citenamefont {Crain},\ and\ \citenamefont
  {Martyna}}]{Jones13a}%
  \BibitemOpen
  \bibfield  {author} {\bibinfo {author} {\bibfnamefont {A.}~\bibnamefont
  {Jones}}, \bibinfo {author} {\bibfnamefont {F.}~\bibnamefont {Cipcigan}},
  \bibinfo {author} {\bibfnamefont {V.~P.}\ \bibnamefont {Sokhan}}, \bibinfo
  {author} {\bibfnamefont {J.}~\bibnamefont {Crain}}, \ and\ \bibinfo {author}
  {\bibfnamefont {G.~J.}\ \bibnamefont {Martyna}},\ }\href {\doibase
  10.1103/PhysRevLett.110.227801} {\bibfield  {journal} {\bibinfo  {journal}
  {Phys. Rev. Lett.}\ }\textbf {\bibinfo {volume} {110}},\ \bibinfo {pages}
  {227801} (\bibinfo {year} {2013}{\natexlab{b}})}\BibitemShut {NoStop}%
\bibitem [{Sup(2015)}]{SuppMat}%
  \BibitemOpen
  \href@noop {} {} (\bibinfo {year} {2015}),\ \bibinfo {note} {see Supplemental
  Material at [\emph{URL will be inserted by publisher}] for the Drude model
  parameters and simulation details}\BibitemShut {NoStop}%
\bibitem [{\citenamefont {Wagner}\ and\ \citenamefont
  {Pru{\ss}}(2002)}]{Wagn02}%
  \BibitemOpen
  \bibfield  {author} {\bibinfo {author} {\bibfnamefont {W.}~\bibnamefont
  {Wagner}}\ and\ \bibinfo {author} {\bibfnamefont {A.}~\bibnamefont
  {Pru{\ss}}},\ }\href {\doibase 10.1063/1.1461829} {\bibfield  {journal}
  {\bibinfo  {journal} {J.\ Phys.\ Chem.\ Ref.\ Data}\ }\textbf {\bibinfo
  {volume} {31}},\ \bibinfo {pages} {387} (\bibinfo {year} {2002})}\BibitemShut
  {NoStop}%
\bibitem [{\citenamefont {Bernal}\ and\ \citenamefont {Fowler}(1933)}]{Bern33}%
  \BibitemOpen
  \bibfield  {author} {\bibinfo {author} {\bibfnamefont {J.~D.}\ \bibnamefont
  {Bernal}}\ and\ \bibinfo {author} {\bibfnamefont {R.~H.}\ \bibnamefont
  {Fowler}},\ }\href {\doibase 10.1063/1.1749327} {\bibfield  {journal}
  {\bibinfo  {journal} {J.\ Chem.\ Phys.}\ }\textbf {\bibinfo {volume} {1}},\
  \bibinfo {pages} {515} (\bibinfo {year} {1933})}\BibitemShut {NoStop}%
\bibitem [{\citenamefont {Hoffmann}\ and\ \citenamefont
  {Conradi}(1997)}]{Hoff97}%
  \BibitemOpen
  \bibfield  {author} {\bibinfo {author} {\bibfnamefont {M.~M.}\ \bibnamefont
  {Hoffmann}}\ and\ \bibinfo {author} {\bibfnamefont {M.~S.}\ \bibnamefont
  {Conradi}},\ }\href {\doibase 10.1021/ja964331g} {\bibfield  {journal}
  {\bibinfo  {journal} {J. Am. Chem. Soc.}\ }\textbf {\bibinfo {volume}
  {119}},\ \bibinfo {pages} {3811} (\bibinfo {year} {1997})}\BibitemShut
  {NoStop}%
\bibitem [{\citenamefont {Wernet}\ \emph {et~al.}(2005)\citenamefont {Wernet},
  \citenamefont {Testemale}, \citenamefont {Hazemann}, \citenamefont {Argoud},
  \citenamefont {Glatzel}, \citenamefont {Pettersson}, \citenamefont
  {Nilsson},\ and\ \citenamefont {Bergmann}}]{Wern05}%
  \BibitemOpen
  \bibfield  {author} {\bibinfo {author} {\bibfnamefont {P.}~\bibnamefont
  {Wernet}}, \bibinfo {author} {\bibfnamefont {D.}~\bibnamefont {Testemale}},
  \bibinfo {author} {\bibfnamefont {J.-L.}\ \bibnamefont {Hazemann}}, \bibinfo
  {author} {\bibfnamefont {R.}~\bibnamefont {Argoud}}, \bibinfo {author}
  {\bibfnamefont {P.}~\bibnamefont {Glatzel}}, \bibinfo {author} {\bibfnamefont
  {L.~G.~M.}\ \bibnamefont {Pettersson}}, \bibinfo {author} {\bibfnamefont
  {A.}~\bibnamefont {Nilsson}}, \ and\ \bibinfo {author} {\bibfnamefont
  {U.}~\bibnamefont {Bergmann}},\ }\href {\doibase 10.1063/1.2064867}
  {\bibfield  {journal} {\bibinfo  {journal} {J.\ Chem.\ Phys.}\ }\textbf
  {\bibinfo {volume} {123}},\ \bibinfo {eid} {154503} (\bibinfo {year}
  {2005})}\BibitemShut {NoStop}%
\bibitem [{\citenamefont {Bernabei}\ \emph {et~al.}(2008)\citenamefont
  {Bernabei}, \citenamefont {Botti}, \citenamefont {Bruni}, \citenamefont
  {Ricci},\ and\ \citenamefont {Soper}}]{Bern08a}%
  \BibitemOpen
  \bibfield  {author} {\bibinfo {author} {\bibfnamefont {M.}~\bibnamefont
  {Bernabei}}, \bibinfo {author} {\bibfnamefont {A.}~\bibnamefont {Botti}},
  \bibinfo {author} {\bibfnamefont {F.}~\bibnamefont {Bruni}}, \bibinfo
  {author} {\bibfnamefont {M.~A.}\ \bibnamefont {Ricci}}, \ and\ \bibinfo
  {author} {\bibfnamefont {A.~K.}\ \bibnamefont {Soper}},\ }\href {\doibase
  10.1103/PhysRevE.78.021505} {\bibfield  {journal} {\bibinfo  {journal}
  {Phys.\ Rev.\ E}\ }\textbf {\bibinfo {volume} {78}},\ \bibinfo {pages}
  {021505} (\bibinfo {year} {2008})}\BibitemShut {NoStop}%
\bibitem [{\citenamefont {Wernet}\ \emph {et~al.}(2004)\citenamefont {Wernet},
  \citenamefont {Nordlund}, \citenamefont {Bergmann}, \citenamefont
  {Cavalleri}, \citenamefont {Odelius}, \citenamefont {Ogasawara},
  \citenamefont {N{\"a}slund}, \citenamefont {Hirsch}, \citenamefont
  {Ojam{\"a}e}, \citenamefont {Glatzel}, \citenamefont {Pettersson},\ and\
  \citenamefont {Nilsson}}]{Wern04}%
  \BibitemOpen
  \bibfield  {author} {\bibinfo {author} {\bibfnamefont {P.}~\bibnamefont
  {Wernet}}, \bibinfo {author} {\bibfnamefont {D.}~\bibnamefont {Nordlund}},
  \bibinfo {author} {\bibfnamefont {U.}~\bibnamefont {Bergmann}}, \bibinfo
  {author} {\bibfnamefont {M.}~\bibnamefont {Cavalleri}}, \bibinfo {author}
  {\bibfnamefont {M.}~\bibnamefont {Odelius}}, \bibinfo {author} {\bibfnamefont
  {H.}~\bibnamefont {Ogasawara}}, \bibinfo {author} {\bibfnamefont {L.~{\AA}.}\
  \bibnamefont {N{\"a}slund}}, \bibinfo {author} {\bibfnamefont {T.~K.}\
  \bibnamefont {Hirsch}}, \bibinfo {author} {\bibfnamefont {L.}~\bibnamefont
  {Ojam{\"a}e}}, \bibinfo {author} {\bibfnamefont {P.}~\bibnamefont {Glatzel}},
  \bibinfo {author} {\bibfnamefont {L.~G.~M.}\ \bibnamefont {Pettersson}}, \
  and\ \bibinfo {author} {\bibfnamefont {A.}~\bibnamefont {Nilsson}},\ }\href
  {\doibase 10.1126/science.1096205} {\bibfield  {journal} {\bibinfo  {journal}
  {Science}\ }\textbf {\bibinfo {volume} {304}},\ \bibinfo {pages} {995}
  (\bibinfo {year} {2004})}\BibitemShut {NoStop}%
\bibitem [{\citenamefont {Head-Gordon}\ and\ \citenamefont
  {Johnson}(2006)}]{Head06}%
  \BibitemOpen
  \bibfield  {author} {\bibinfo {author} {\bibfnamefont {T.}~\bibnamefont
  {Head-Gordon}}\ and\ \bibinfo {author} {\bibfnamefont {M.~E.}\ \bibnamefont
  {Johnson}},\ }\href {\doibase 10.1073/pnas.0510593103} {\bibfield  {journal}
  {\bibinfo  {journal} {Proc.\ Natl Acad.\ Sci.\ USA}\ }\textbf {\bibinfo
  {volume} {103}},\ \bibinfo {pages} {7973} (\bibinfo {year}
  {2006})}\BibitemShut {NoStop}%
\bibitem [{\citenamefont {Sun}\ \emph {et~al.}(2014)\citenamefont {Sun},
  \citenamefont {Wang},\ and\ \citenamefont {Ding}}]{Sun14}%
  \BibitemOpen
  \bibfield  {author} {\bibinfo {author} {\bibfnamefont {Q.}~\bibnamefont
  {Sun}}, \bibinfo {author} {\bibfnamefont {Q.}~\bibnamefont {Wang}}, \ and\
  \bibinfo {author} {\bibfnamefont {D.}~\bibnamefont {Ding}},\ }\href {\doibase
  10.1021/jp503474s} {\bibfield  {journal} {\bibinfo  {journal} {J.\ Phys.\
  Chem.\ B}\ }\textbf {\bibinfo {volume} {118}},\ \bibinfo {pages} {11253}
  (\bibinfo {year} {2014})}\BibitemShut {NoStop}%
\bibitem [{\citenamefont {Kumar}\ \emph {et~al.}(2007)\citenamefont {Kumar},
  \citenamefont {Schmidt},\ and\ \citenamefont {Skinner}}]{Kuma07}%
  \BibitemOpen
  \bibfield  {author} {\bibinfo {author} {\bibfnamefont {R.}~\bibnamefont
  {Kumar}}, \bibinfo {author} {\bibfnamefont {J.~R.}\ \bibnamefont {Schmidt}},
  \ and\ \bibinfo {author} {\bibfnamefont {J.~L.}\ \bibnamefont {Skinner}},\
  }\href {\doibase 10.1063/1.2742385} {\bibfield  {journal} {\bibinfo
  {journal} {J.\ Chem.\ Phys.}\ }\textbf {\bibinfo {volume} {126}},\ \bibinfo
  {pages} {204107} (\bibinfo {year} {2007})}\BibitemShut {NoStop}%
\bibitem [{\citenamefont {Blumberg}\ \emph {et~al.}(1984)\citenamefont
  {Blumberg}, \citenamefont {Stanley}, \citenamefont {Geiger},\ and\
  \citenamefont {Mausbach}}]{Blum84}%
  \BibitemOpen
  \bibfield  {author} {\bibinfo {author} {\bibfnamefont {R.~L.}\ \bibnamefont
  {Blumberg}}, \bibinfo {author} {\bibfnamefont {H.~E.}\ \bibnamefont
  {Stanley}}, \bibinfo {author} {\bibfnamefont {A.}~\bibnamefont {Geiger}}, \
  and\ \bibinfo {author} {\bibfnamefont {P.}~\bibnamefont {Mausbach}},\ }\href
  {\doibase http://dx.doi.org/10.1063/1.446593} {\bibfield  {journal} {\bibinfo
   {journal} {J.\ Chem.\ Phys.}\ }\textbf {\bibinfo {volume} {80}},\ \bibinfo
  {pages} {5230} (\bibinfo {year} {1984})}\BibitemShut {NoStop}%
\bibitem [{\citenamefont {Soper}(2000)}]{Soper00}%
  \BibitemOpen
  \bibfield  {author} {\bibinfo {author} {\bibfnamefont {A.~K.}\ \bibnamefont
  {Soper}},\ }\href {\doibase 10.1016/S0301-0104(00)00179-8} {\bibfield
  {journal} {\bibinfo  {journal} {Chem.\ Phys.}\ }\textbf {\bibinfo {volume}
  {258}},\ \bibinfo {pages} {121} (\bibinfo {year} {2000})}\BibitemShut
  {NoStop}%
\bibitem [{\citenamefont {Gregory}\ \emph {et~al.}(1997)\citenamefont
  {Gregory}, \citenamefont {Clary}, \citenamefont {Liu}, \citenamefont
  {Brown},\ and\ \citenamefont {Saykally}}]{Greg97}%
  \BibitemOpen
  \bibfield  {author} {\bibinfo {author} {\bibfnamefont {J.~K.}\ \bibnamefont
  {Gregory}}, \bibinfo {author} {\bibfnamefont {D.~C.}\ \bibnamefont {Clary}},
  \bibinfo {author} {\bibfnamefont {K.}~\bibnamefont {Liu}}, \bibinfo {author}
  {\bibfnamefont {M.~G.}\ \bibnamefont {Brown}}, \ and\ \bibinfo {author}
  {\bibfnamefont {R.~J.}\ \bibnamefont {Saykally}},\ }\href {\doibase
  10.1126/science.275.5301.814} {\bibfield  {journal} {\bibinfo  {journal}
  {Science}\ }\textbf {\bibinfo {volume} {275}},\ \bibinfo {pages} {814}
  (\bibinfo {year} {1997})}\BibitemShut {NoStop}%
\bibitem [{\citenamefont {Onsager}(1936)}]{Onsa36}%
  \BibitemOpen
  \bibfield  {author} {\bibinfo {author} {\bibfnamefont {L.}~\bibnamefont
  {Onsager}},\ }\href {\doibase 10.1021/ja01299a050} {\bibfield  {journal}
  {\bibinfo  {journal} {J.\ Am.\ Chem.\ Soc.}\ }\textbf {\bibinfo {volume}
  {58}},\ \bibinfo {pages} {1486} (\bibinfo {year} {1936})}\BibitemShut
  {NoStop}%
\bibitem [{\citenamefont {Partay}\ \emph {et~al.}(2007)\citenamefont {Partay},
  \citenamefont {Jedlovszky}, \citenamefont {Brovchenko},\ and\ \citenamefont
  {Oleinikova}}]{Part07}%
  \BibitemOpen
  \bibfield  {author} {\bibinfo {author} {\bibfnamefont {L.~B.}\ \bibnamefont
  {Partay}}, \bibinfo {author} {\bibfnamefont {P.}~\bibnamefont {Jedlovszky}},
  \bibinfo {author} {\bibfnamefont {I.}~\bibnamefont {Brovchenko}}, \ and\
  \bibinfo {author} {\bibfnamefont {A.}~\bibnamefont {Oleinikova}},\ }\href
  {\doibase 10.1021/jp070575j} {\bibfield  {journal} {\bibinfo  {journal} {J.\
  Phys.\ Chem.\ B}\ }\textbf {\bibinfo {volume} {111}},\ \bibinfo {pages}
  {7603} (\bibinfo {year} {2007})}\BibitemShut {NoStop}%
\bibitem [{\citenamefont {Lamoureux}\ \emph {et~al.}(2006)\citenamefont
  {Lamoureux}, \citenamefont {Harder}, \citenamefont {Vorobyov}, \citenamefont
  {Roux},\ and\ \citenamefont {Jr.}}]{Lamo06}%
  \BibitemOpen
  \bibfield  {author} {\bibinfo {author} {\bibfnamefont {G.}~\bibnamefont
  {Lamoureux}}, \bibinfo {author} {\bibfnamefont {E.}~\bibnamefont {Harder}},
  \bibinfo {author} {\bibfnamefont {I.~V.}\ \bibnamefont {Vorobyov}}, \bibinfo
  {author} {\bibfnamefont {B.}~\bibnamefont {Roux}}, \ and\ \bibinfo {author}
  {\bibfnamefont {A.~D.~M.}\ \bibnamefont {Jr.}},\ }\href {\doibase
  10.1016/j.cplett.2005.10.135} {\bibfield  {journal} {\bibinfo  {journal}
  {Chem.\ Phys.\ Lett.}\ }\textbf {\bibinfo {volume} {418}},\ \bibinfo {pages}
  {245} (\bibinfo {year} {2006})}\BibitemShut {NoStop}%
\bibitem [{\citenamefont {Kiss}\ and\ \citenamefont {Baranyai}(2012)}]{Kiss12}%
  \BibitemOpen
  \bibfield  {author} {\bibinfo {author} {\bibfnamefont {P.~T.}\ \bibnamefont
  {Kiss}}\ and\ \bibinfo {author} {\bibfnamefont {A.}~\bibnamefont
  {Baranyai}},\ }\href {\doibase 10.1063/1.4767064} {\bibfield  {journal}
  {\bibinfo  {journal} {J.\ Chem.\ Phys.}\ }\textbf {\bibinfo {volume} {137}},\
  \bibinfo {pages} {194103} (\bibinfo {year} {2012})}\BibitemShut {NoStop}%
\bibitem [{\citenamefont {van Leeuwen}(1994)}]{Leeu94}%
  \BibitemOpen
  \bibfield  {author} {\bibinfo {author} {\bibfnamefont {M.}~\bibnamefont {van
  Leeuwen}},\ }\href {\doibase 10.1016/0378-3812(94)80018-9} {\bibfield
  {journal} {\bibinfo  {journal} {Fluid Phase Equilibria}\ }\textbf {\bibinfo
  {volume} {99}},\ \bibinfo {pages} {1} (\bibinfo {year} {1994})}\BibitemShut
  {NoStop}%
\end{thebibliography}%


\begin{thebibliography}{11}
\expandafter\ifx\csname natexlab\endcsname\relax\def\natexlab#1{#1}\fi
\expandafter\ifx\csname bibnamefont\endcsname\relax
  \def\bibnamefont#1{#1}\fi
\expandafter\ifx\csname bibfnamefont\endcsname\relax
  \def\bibfnamefont#1{#1}\fi
\expandafter\ifx\csname citenamefont\endcsname\relax
  \def\citenamefont#1{#1}\fi
\expandafter\ifx\csname url\endcsname\relax
  \def\url#1{\texttt{#1}}\fi
\expandafter\ifx\csname urlprefix\endcsname\relax\def\urlprefix{URL }\fi
\providecommand{\bibinfo}[2]{#2}
\providecommand{\eprint}[2][]{\url{#2}}

\bibitem[{\citenamefont{Jones et~al.}(2013{\natexlab{a}})\citenamefont{Jones,
  Cipcigan, Sokhan, Crain, and Martyna}}]{Jones13a}
\bibinfo{author}{\bibfnamefont{A.}~\bibnamefont{Jones}},
  \bibinfo{author}{\bibfnamefont{F.}~\bibnamefont{Cipcigan}},
  \bibinfo{author}{\bibfnamefont{V.~P.} \bibnamefont{Sokhan}},
  \bibinfo{author}{\bibfnamefont{J.}~\bibnamefont{Crain}}, \bibnamefont{and}
  \bibinfo{author}{\bibfnamefont{G.~J.} \bibnamefont{Martyna}},
  \bibinfo{journal}{Phys. Rev. Lett.} \textbf{\bibinfo{volume}{110}},
  \bibinfo{pages}{227801} (\bibinfo{year}{2013}{\natexlab{a}}).

\bibitem[{\citenamefont{Jones et~al.}(2009)\citenamefont{Jones, Thompson,
  Crain, Muser, and Martyna}}]{Jones09}
\bibinfo{author}{\bibfnamefont{A.}~\bibnamefont{Jones}},
  \bibinfo{author}{\bibfnamefont{A.}~\bibnamefont{Thompson}},
  \bibinfo{author}{\bibfnamefont{J.}~\bibnamefont{Crain}},
  \bibinfo{author}{\bibfnamefont{M.~H.} \bibnamefont{Muser}}, \bibnamefont{and}
  \bibinfo{author}{\bibfnamefont{G.~J.} \bibnamefont{Martyna}},
  \bibinfo{journal}{Phys.\ Rev.\ B} \textbf{\bibinfo{volume}{79}},
  \bibinfo{pages}{144119} (\bibinfo{year}{2009}).

\bibitem[{\citenamefont{Jones et~al.}(2013{\natexlab{b}})\citenamefont{Jones,
  Crain, Sokhan, Whitfield, and Martyna}}]{Jones13}
\bibinfo{author}{\bibfnamefont{A.~P.} \bibnamefont{Jones}},
  \bibinfo{author}{\bibfnamefont{J.}~\bibnamefont{Crain}},
  \bibinfo{author}{\bibfnamefont{V.~P.} \bibnamefont{Sokhan}},
  \bibinfo{author}{\bibfnamefont{T.~W.} \bibnamefont{Whitfield}},
  \bibnamefont{and} \bibinfo{author}{\bibfnamefont{G.~J.}
  \bibnamefont{Martyna}}, \bibinfo{journal}{Phys.\ Rev.\ B}
  \textbf{\bibinfo{volume}{87}}, \bibinfo{pages}{144103}
  (\bibinfo{year}{2013}{\natexlab{b}}).

\bibitem[{\citenamefont{Martyna et~al.}(1996)\citenamefont{Martyna, Tuckerman,
  Tobias, and Klein}}]{Martyna96}
\bibinfo{author}{\bibfnamefont{G.~J.} \bibnamefont{Martyna}},
  \bibinfo{author}{\bibfnamefont{M.~E.} \bibnamefont{Tuckerman}},
  \bibinfo{author}{\bibfnamefont{D.~J.} \bibnamefont{Tobias}},
  \bibnamefont{and} \bibinfo{author}{\bibfnamefont{M.~L.} \bibnamefont{Klein}},
  \bibinfo{journal}{Mol.\ Phys.} \textbf{\bibinfo{volume}{87}},
  \bibinfo{pages}{1117} (\bibinfo{year}{1996}).

\bibitem[{\citenamefont{Whitfield and Martyna}(2007)}]{Whit07}
\bibinfo{author}{\bibfnamefont{T.~W.} \bibnamefont{Whitfield}}
  \bibnamefont{and} \bibinfo{author}{\bibfnamefont{G.~J.}
  \bibnamefont{Martyna}}, \bibinfo{journal}{J.\ Chem.\ Phys.}
  \textbf{\bibinfo{volume}{126}}, \bibinfo{eid}{074104} (\bibinfo{year}{2007}).

\bibitem[{\citenamefont{Jones et~al.}(2013{\natexlab{c}})\citenamefont{Jones,
  Crain, Cipcigan, Sokhan, Modani, and Martyna}}]{Jones13b}
\bibinfo{author}{\bibfnamefont{A.}~\bibnamefont{Jones}},
  \bibinfo{author}{\bibfnamefont{J.}~\bibnamefont{Crain}},
  \bibinfo{author}{\bibfnamefont{F.}~\bibnamefont{Cipcigan}},
  \bibinfo{author}{\bibfnamefont{V.}~\bibnamefont{Sokhan}},
  \bibinfo{author}{\bibfnamefont{M.}~\bibnamefont{Modani}}, \bibnamefont{and}
  \bibinfo{author}{\bibfnamefont{G.}~\bibnamefont{Martyna}},
  \bibinfo{journal}{Mol.\ Phys.} \textbf{\bibinfo{volume}{111}},
  \bibinfo{pages}{3465} (\bibinfo{year}{2013}{\natexlab{c}}).

\bibitem[{\citenamefont{Jones and Leimkuhler}(2011)}]{Jones11}
\bibinfo{author}{\bibfnamefont{A.}~\bibnamefont{Jones}} \bibnamefont{and}
  \bibinfo{author}{\bibfnamefont{B.}~\bibnamefont{Leimkuhler}},
  \bibinfo{journal}{J.\ Chem.\ Phys.} \textbf{\bibinfo{volume}{135}},
  \bibinfo{pages}{084125} (\bibinfo{year}{2011}).

\bibitem[{\citenamefont{Kumar et~al.}(2007)\citenamefont{Kumar, Schmidt, and
  Skinner}}]{Kuma07}
\bibinfo{author}{\bibfnamefont{R.}~\bibnamefont{Kumar}},
  \bibinfo{author}{\bibfnamefont{J.~R.} \bibnamefont{Schmidt}},
  \bibnamefont{and} \bibinfo{author}{\bibfnamefont{J.~L.}
  \bibnamefont{Skinner}}, \bibinfo{journal}{J.\ Chem.\ Phys.}
  \textbf{\bibinfo{volume}{126}}, \bibinfo{pages}{204107}
  (\bibinfo{year}{2007}).

\bibitem[{\citenamefont{Onsager}(1936)}]{Onsa36}
\bibinfo{author}{\bibfnamefont{L.}~\bibnamefont{Onsager}},
  \bibinfo{journal}{J.\ Am.\ Chem.\ Soc.} \textbf{\bibinfo{volume}{58}},
  \bibinfo{pages}{1486} (\bibinfo{year}{1936}).

\bibitem[{\citenamefont{van Leeuwen}(1994)}]{Leeu94}
\bibinfo{author}{\bibfnamefont{M.}~\bibnamefont{van Leeuwen}},
  \bibinfo{journal}{Fluid Phase Equilibria} \textbf{\bibinfo{volume}{99}},
  \bibinfo{pages}{1} (\bibinfo{year}{1994}).

\bibitem[{\citenamefont{Phillips et~al.}(2005)\citenamefont{Phillips, Braun,
  Wang, Gumbart, Tajkhorshid, Villa, Chipot, Skeel, Kalé, and
  Schulten}}]{NAMD}
\bibinfo{author}{\bibfnamefont{J.~C.} \bibnamefont{Phillips}},
  \bibinfo{author}{\bibfnamefont{R.}~\bibnamefont{Braun}},
  \bibinfo{author}{\bibfnamefont{W.}~\bibnamefont{Wang}},
  \bibinfo{author}{\bibfnamefont{J.}~\bibnamefont{Gumbart}},
  \bibinfo{author}{\bibfnamefont{E.}~\bibnamefont{Tajkhorshid}},
  \bibinfo{author}{\bibfnamefont{E.}~\bibnamefont{Villa}},
  \bibinfo{author}{\bibfnamefont{C.}~\bibnamefont{Chipot}},
  \bibinfo{author}{\bibfnamefont{R.~D.} \bibnamefont{Skeel}},
  \bibinfo{author}{\bibfnamefont{L.}~\bibnamefont{Kalé}}, \bibnamefont{and}
  \bibinfo{author}{\bibfnamefont{K.}~\bibnamefont{Schulten}},
  \bibinfo{journal}{J.\ Comp.\ Chem.} \textbf{\bibinfo{volume}{26}},
  \bibinfo{pages}{1781} (\bibinfo{year}{2005}).

\end{thebibliography}
%--------------------------------------------------------------------------------
%\clearpage

%\afterpage{%

\clearpage
%}
\end{document}

\widetext
\title{Supplemental Material: \\ ~~ \\
 Molecular-scale signatures of the liquid-gas transition observed in supercritical water}

\author{V.~P.~Sokhan}
\affiliation{National Physical Laboratory, Hampton Road, Teddington, Middlesex TW11 0LW, UK}

\author{A.~Jones}
\affiliation{School of Physics and Astronomy, The University of Edinburgh,
             Mayfield Road, Edinburgh EH9 3JZ, UK}

\author{F.~S.~Cipcigan}
\affiliation{National Physical Laboratory, Hampton Road, Teddington, Middlesex TW11 0LW, UK}
\affiliation{School of Physics and Astronomy, The University of Edinburgh,
             Mayfield Road, Edinburgh EH9 3JZ, UK}

\author{J.~Crain}
\affiliation{National Physical Laboratory, Hampton Road, Teddington, Middlesex TW11 0LW, UK}
\affiliation{School of Physics and Astronomy, The University of Edinburgh,
             Mayfield Road, Edinburgh EH9 3JZ, UK}

\author{G.~J.~Martyna}
\affiliation{IBM T.~J.~Watson Research Center, Yorktown Heights, New York 10598, USA}

\maketitle
%\setstretch{1.5}
\begin{spacing}{1.5}

\subsection{The layout}
A description of the model and simulation methods are provided first. Second, the hydrogen bond definition employed in the text is described and justified. A mean-field treatment of the 
scaling of molecular moment with density is given next. The Stockmayer model is described next followed by a discussion of finite size effects. Last, we address the different local association topology of water models and the Stockmayer fluid.

\subsection{The model and simulation details}
\begin{figure}[H]    %----------------------------------------- Figure 1 ------
 \centering
 \includegraphics{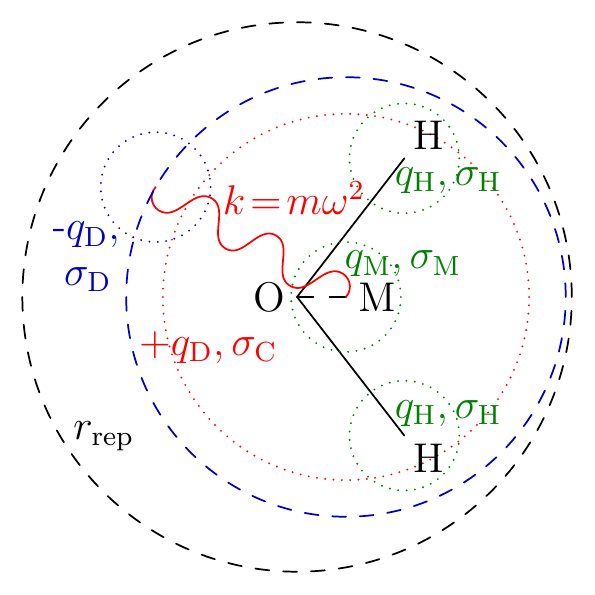}
 \caption{QDO model for water.}\label{Fig:H2O}
\end{figure}

For water, which has nearly isotropic polarizability, we follow the model 
construction reported in Ref.~\cite{Jones13a}.
The quantum Drude oscillator (QDO) model for water, sketched in Fig.~\ref{Fig:H2O}, 
is based on rigid TIP4P geometry of the monomer ($r_\text{OH}=95.72$~pm, 
$\angle\text{HOH}=104.52^\circ$). 
A \emph{drudon} (quantum Drude oscillator) is tethered to a point M
along the $\angle\text{HOH}$ bisector at a distance $r_\text{OM}=26.67$~pm from 
the oxygen nucleus. 
The model parameters are chosen such that the long-range responses 
match reference values of the monomer and dimer (gas-phase polarizabilities 
and pair-dispersion coefficients). Three fixed point charges, two on hydrogens,
$q_\text{H}=0.605|e|$, and one on M--point, $q_\text{M}=-2q_\text{H}$,
are taken to represent the isolated molecule limit, matched to the low order 
electrostatic moments. The drudon is also tethered to M--point. Its parameters 
in atomic units are: mass, $m_\text{D}=0.3656\,m_e$, charge, $q_\text{D}=-1.1973|e|$, 
and angular frequency, $\omega=0.6287\,E_h/\hbar$. 

The Coulomb forces are regularized at short range by assigning small Gaussian 
widths to the charges, $\sigma_\text{H}=\sigma_\text{D}=0.1a_0$, and 
$\sigma_\text{C}=1.2a_0$ for the QDO center. To correct for missing in the QDO 
model Fermion exchange a biexponential term
\begin{equation*}
\phi(r) = \kappa_1\exp(-\lambda_1 r) + \kappa_2\exp(-\lambda_2 r)
\end{equation*}
has been added between oxygens, providing the required repulsion. The 
parameters $\kappa=\{613.3,10.5693\}$ 
and $\lambda=\{2.3244,1.5145\}$ were obtained from a fit to the difference 
between a reference quantum chemical dimer ground state potential energy 
surface (CCSD(T)-level) and that of the QDO's, calculated using norm-conserving 
diffusion Monte Carlo \cite{Jones09}. In order to reproduce accurately the 
equation of state fully convergent ab initio calculations required for the 
reference potential surface. In Mark I model we fit to empirical potential 
\cite{Jones13a}, which effectively included manybody interactions, and as 
a result, the experimental density was reproduced when dispersion 
interactions were scaled up by 37\%.

The system is thus defined by monomers with ``perfect'' (by construction) 
long-range electrostatics and all resulting condensed phase behavior emerges 
as a model prediction. No additional thermodynamic data or liquid state 
information was used.

The coarse grained Hamiltonian for the quantum Drude oscillator (QDO) water model is
\begin{gather}
H = \sum_iT_i^{(\text{rig})} + \phi^{(\text{Coul})}(\mathbf{R}) 
    + \sum_{j\neq i}\phi^{(\text{rep})} \left(R_{\text{O}ij}\right) 
    + E_0^{(\text{D})}(\mathbf{R}), \nonumber \\
\hat{\mathbf{H}}^{(\text{D})}\psi_0(\mathbf{r},\mathbf{R}) 
= E_0^{(\text{D})}(\mathbf{R})\,\psi_0(\mathbf{r},\mathbf{R}), \nonumber \\
\hat{\mathbf{H}}^{(\text{D})} = \sum_i\left(\hat{\mathbf{T}}_i 
+ \frac{\mu\omega^2}{2}(\mathbf{r}_i-\mathbf{R}_{\text{c}i})^2\right) 
+ \phi^{(\text{Coul})}(\mathbf{r},\mathbf{R}). \nonumber
\end{gather}
Here, $T_i^{(\text{rig})}$ is the classical rigid body kinetic energy of 
water molecule $i$; $9N$-vector $\mathbf{R}$ represents the coordinates of 
all the water molecules in the system; $\phi^{(\text{Coul})}(\mathbf{R})$ 
is the intermolecular Coulomb interaction energy between the fixed monomer 
gas phase charge distributions only; $\phi^{(\text{rep})}\left(R_{\text{O}ij}\right)$ 
is an oxygen-centered pair-wise repulsion. The quantities $E_0^{(\text{D})}(\mathbf{R})$ 
and $\psi_0(\mathbf{r},\mathbf{R})$ are the ground state Born--Oppenheimer energy 
surface and wavefunction of the QDO electronic structure, respectively; $\mathbf{r}_i$
and $\hat{\mathbf{T}}_i$ are the position and quantum kinetic energy of the oscillator 
centered on molecule $i$. The oscillators centered at $\mathbf{R}_{\text{c}i}$ are
characterized by parameters $\{\mu,\omega,q\}$. Finally, 
$\phi^{(\text{Coul})}(\mathbf{r},\mathbf{R})$ is intermolecular Coulomb interaction 
between the Drude particles of charge $q^-$, at position $\mathbf{r}$, their 
center of oscillation with charge $q^+$, at position $\mathbf{R}_{\text{c}}$, 
and the fixed monomer charge distribution. Regularization of the Coulomb 
interaction is discussed elsewhere \cite{Jones13}.

Finite temperature condensed phase calculations were performed using adiabatic 
path integral molecular dynamics for quantum Drude oscillators (APIMD-QDO) 
\cite{Martyna96,Whit07,Jones13b} in the canonical and isothermal-isobaric 
ensembles \cite{Jones11}. We used the discretized (high-temperature) path 
integral approximation to the partition function with the number of beads
(the Trotter number) $P=96$ and adiabaticity factor $\gamma=16$ and 
close to Widom line densities, $\gamma=32$.

All calculations were performed for canonical ($NVT$) or isothermal-isobaric 
($NpT$) conditions using cubic 
systems containing 300 water molecules with 3D periodic boundary conditions 
and Ewald summation technique for electrostatic interactions. 
At 673~K we recalculated all density points with $N=1000$ molecules and 
the results were statistically identical to that smaller system. This is the 
consequence of the absence of any truncation scheme in intermolecular 
interactions and of the short range of structural and orientational ordering 
at high temperatures. The integration timestep for adiabaticity factor 
$\gamma=16$ was $dt=0.15$~fs ($6.25\,\hbar/E_h$) and a typical integration time 
was 150~ps ($10^6dt$), which requires ca.~27~h of CPU time on BlueGene/Q 
running on 8K cores.

\subsection{Finite size effects estimation}

\begin{figure}[H]    %----------------------------------------- Figure 2 ------
 \centering
 \includegraphics{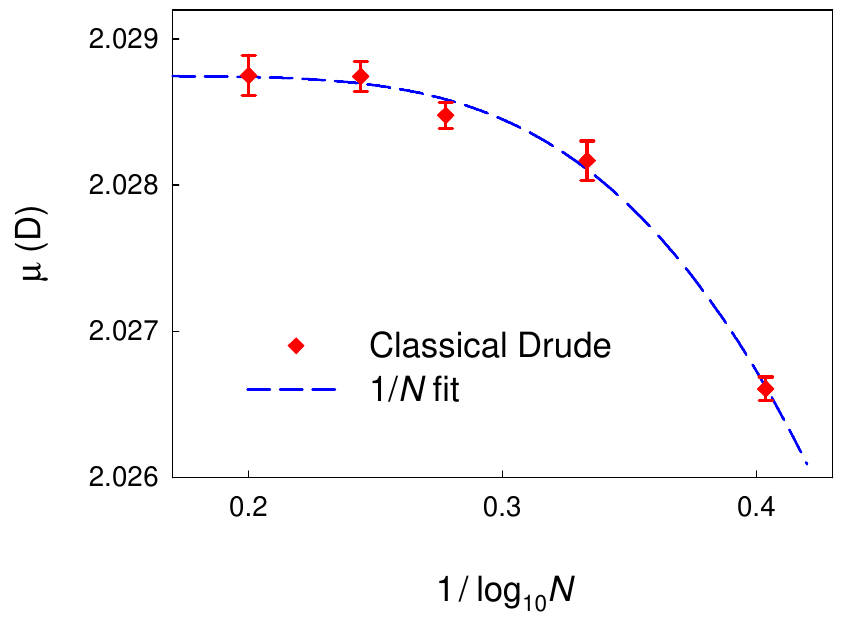}
 \caption{System size dependence of the molecular dipole moment calculated for 
          the classical Drude model at reduced density $\rho^*=0.82$
					and reduced temperature $T^*=1.036$. The error bars correspond to 
					two standard deviations (95\% confidence level).}\label{Fig:Syssize}
\end{figure}

In order to verify that our results converged with respect to integration time 
and system size, for one thermodynamic point, $T^*=1.036$ and density $\rho^*=0.82$ 
we performed a set of test calculations using SWM4-NDP model for systems with 
$N=300$, 1,000, 4,000, 12,500, and 100,000 molecules. The Drude oscillator degrees 
of freedom were treated as dynamic variabled and extended Lagrangian equations of 
motion were integrated with 0.5~fs timestep, and the temperature was controlled 
by two thermostats with classical Drude oscillators kept close to 1~K by Langevin 
thermostat. The obtained results for the dipole moment convergence are shown in 
Fig.~\ref{Fig:Syssize}.
They fit well to $1/N$ (shown in the Figure by the dashed line) and from this 
fit the dipole moment extrapolated to infinite size is only than 0.1\% higher 
than the value obtained with the smallest studied system. No systematic 
drift in statistical properties has been observed after equilibrating the 
systems for at least 200~ps.

Special care was taken to ensure that there are no systematic drift in computer 
properties, for which long runs, from 0.1~\si{\micro}s for $N=300$ to 1~ns for 
$N=10^5$ have been performed. In all cases pressure was $p=24.81(4)$~MPa.

\subsection{Hydrogen bonding}
\begin{figure}[t!]    %----------------------------------------- Figure 3 ------
 \begin{minipage}[t]{1.0\textwidth}
	\centering
	\includegraphics[width=80mm]{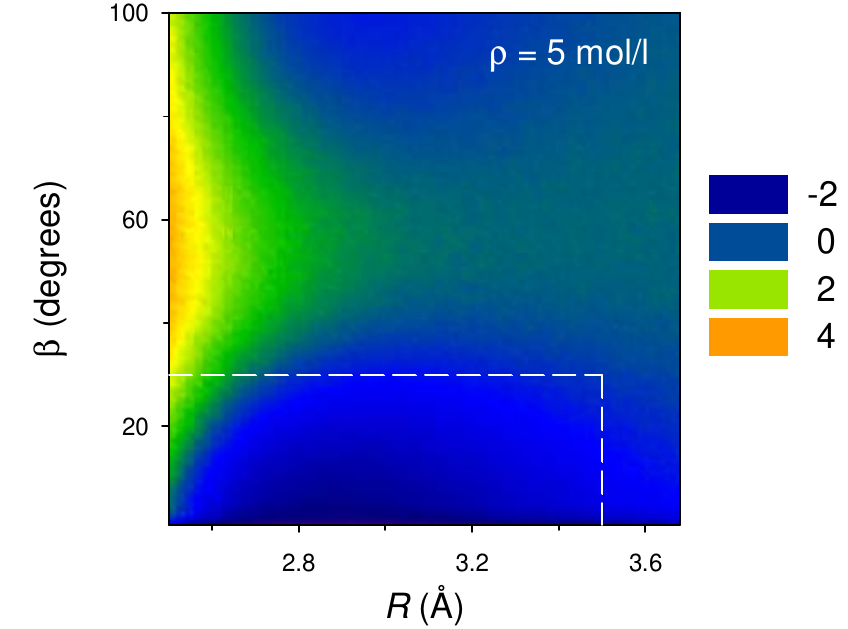}\includegraphics[width=80mm]{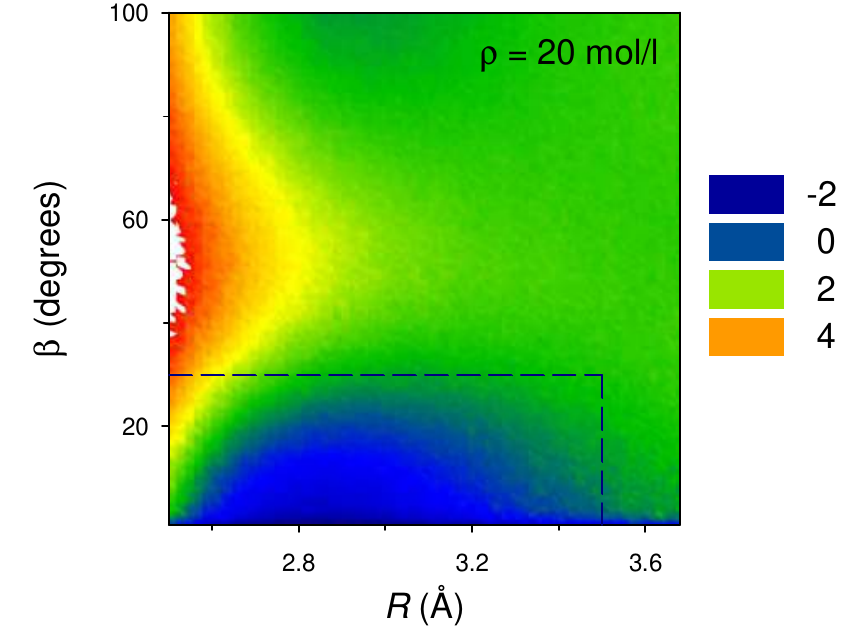}
 {~}
 \end{minipage}

 \begin{minipage}[t]{1.0\textwidth}
	\centering
  \includegraphics[width=80mm]{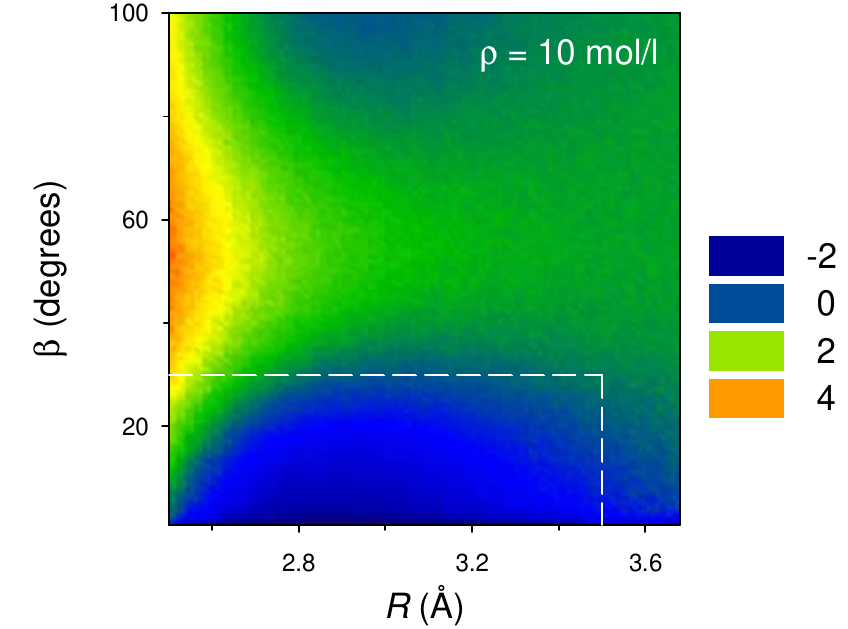}\includegraphics[width=80mm]{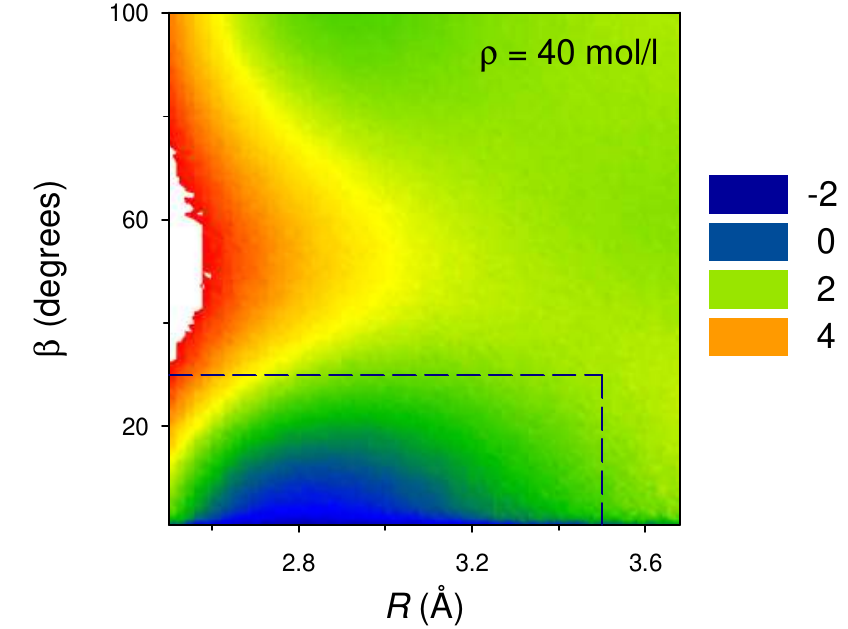}
  {~}
 \end{minipage}

 \begin{minipage}[t]{1.0\textwidth}
	\centering
  \includegraphics[width=80mm]{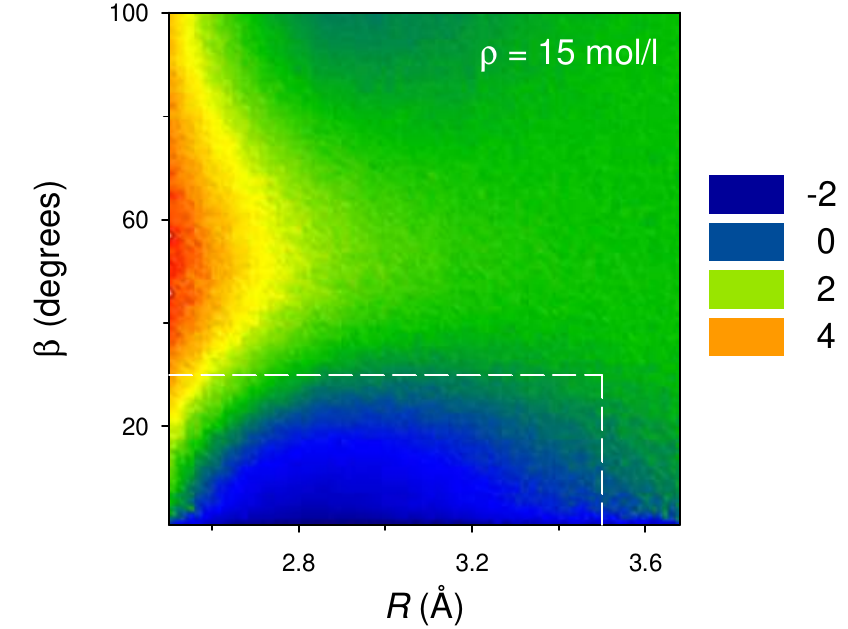}\includegraphics[width=80mm]{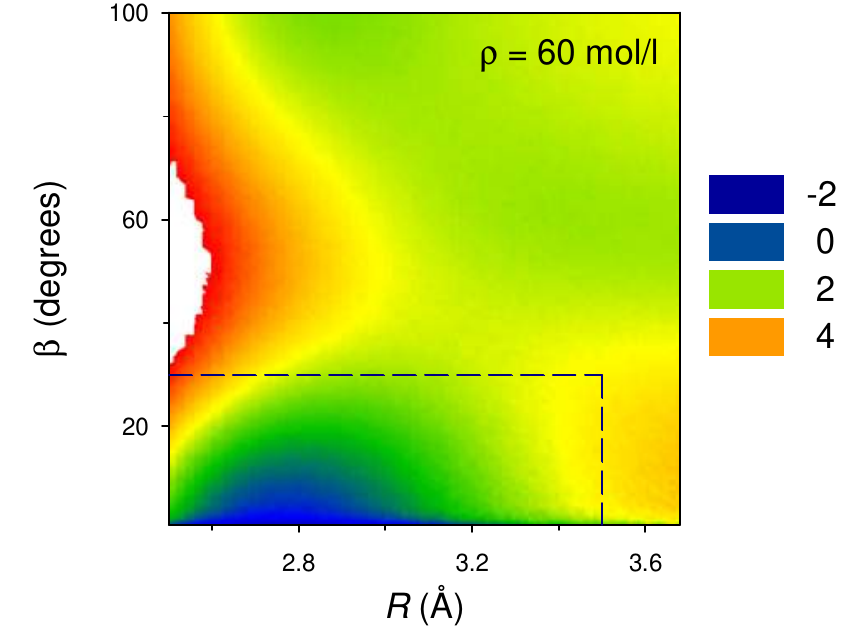}
 \end{minipage}
  \caption{Potential of mean force (in $kT$ units) map in O--O and 
	$\protect\angle\text{OH}\cdots\text{O}$ dimensions at various densities along 
	 the 673~K isotherm. The dashed lines of contrast colors fence the area of 
	 hydrogen-bonded molecules using our criterion.}
	\label{Figs:PMF}
\end{figure}

When calculating various HB distributions we used the geometric $R-\beta$ 
distance-angle definition of the hydrogen bond \cite{Kuma07}, in which two 
water molecules are considered hydrogen-bonded if and only if the O--O 
distance between two molecules is shorter than the certain threshold value, 
$R_\text{HB}$, and the angle $\beta$ between O--H bond and the candidate 
molecule oxygen, $\text{O}-\text{H}\cdots\text{O}$, is smaller than the 
specified criterion $\beta_\text{HB}$. We have found that the values 
$R_\text{HB}=3.5$~\AA{} and $\beta_\text{HB}=30^\circ$ are suitable for 
the studied cases (see Fig.~\ref{Figs:PMF}).

\subsection{Molecular moment scaling}
Molecular dipole moment scales linearly with density. This could be understood 
from simple argument.

The second contribution can be treated in mean-field approximation. 
By the chain rule for differentiation,
\begin{equation}
  \dfrac{\partial|\bar{\mu}|}{\partial\rho} = \dfrac{\partial|\bar{\mu}|}
	  {\partial E_l}\dfrac{\partial E_l}{\partial R}\dfrac{\partial R}{\partial\rho}.
		\label{eq:dip}
\end{equation}
Here, $E_l$ is the projection of the local Coulomb field on the molecule along the 
$\angle$HOH bisector, and $R$ is a measure of average molecular separation.
The first factor on the RHS of (\ref{eq:dip}) is the molecular 
polarizability $\alpha$ along the field direction, to leading order. 
Defining the density as $\rho = mR^{-3}$ we have $\partial R/\partial\rho\sim -R^4/m$. 
For the local field dependence on $R$, we assume leading term should be 
due to the permanent dipole $\mu_0$, the separation $R$ and an orientational 
factor $f(\theta)$ according to $\mu_0 f(\theta)/R^3$. 
Therefore $({\partial E}/{\partial R}) \sim -\mu_0f(\theta)/R^4$ 
and, finally, ${\partial \mu}/{\partial \rho} \sim \alpha\, \mu_0f(\theta)/m$. 
Consequently, we expect the molecular dipole to depend approximately linearly 
on density with the observed change of slope indicative of an increase in 
local density as the HB network coalesces. This is consistent with the 
density dependence of the Onsager reaction field \cite{Onsa36}.

\begin{figure}[t!]    %----------------------------------------- Figure 4 ------
 \begin{minipage}[t]{1.0\textwidth}
	\centering
	\includegraphics[width=68mm]{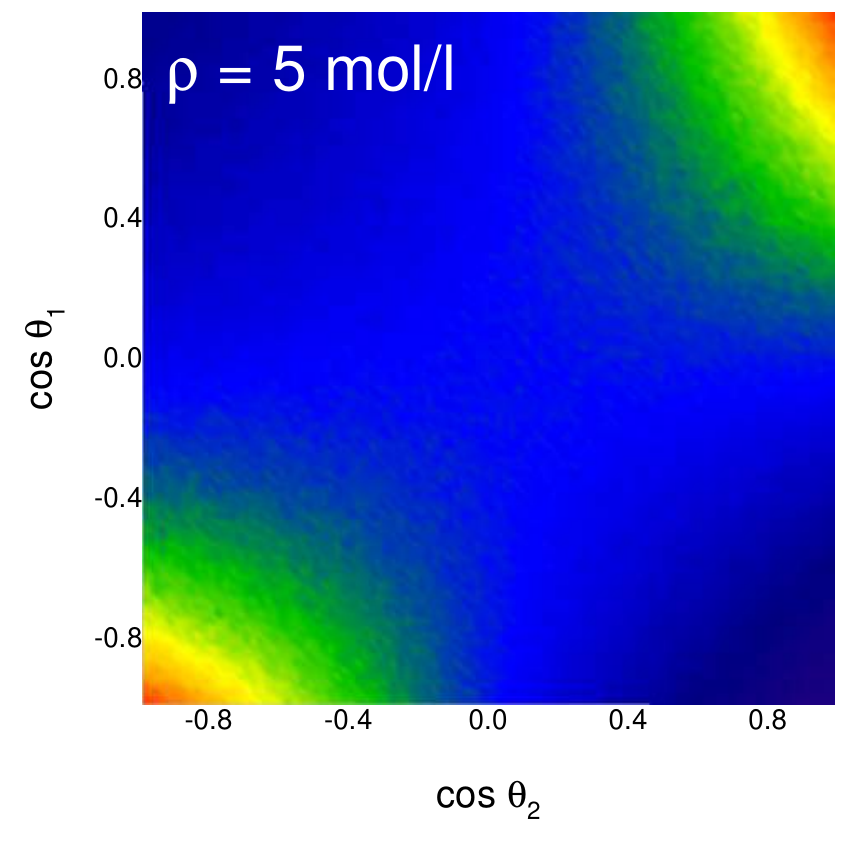}\hfill\includegraphics[width=68mm]{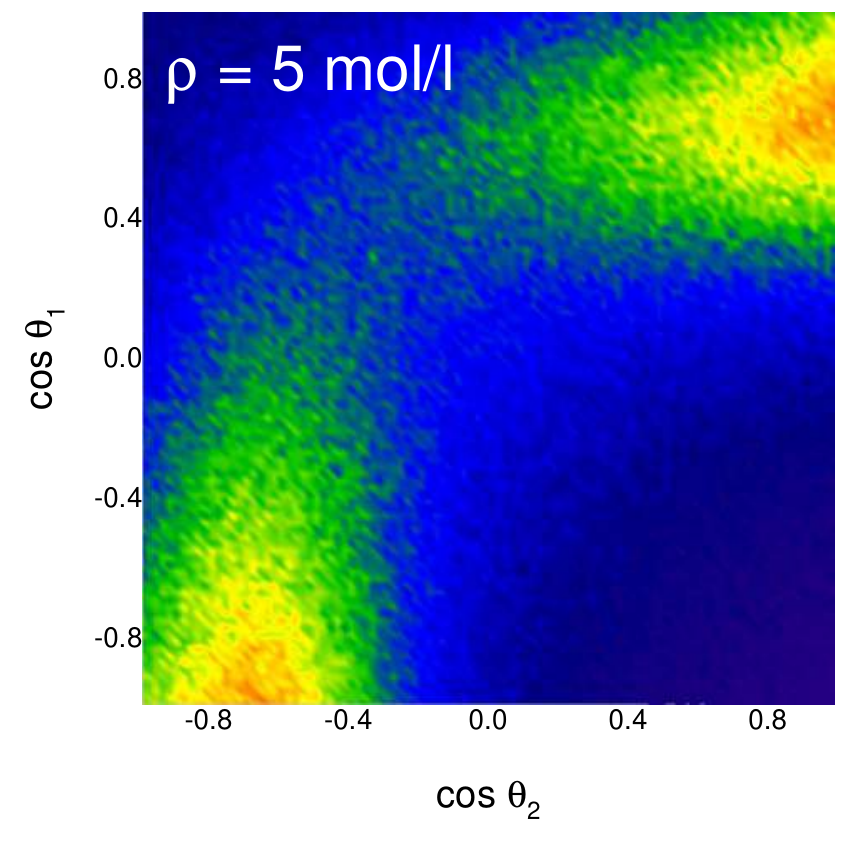}
 {~}
 \end{minipage}

 \begin{minipage}[t]{1.0\textwidth}
	\centering
  \includegraphics[width=68mm]{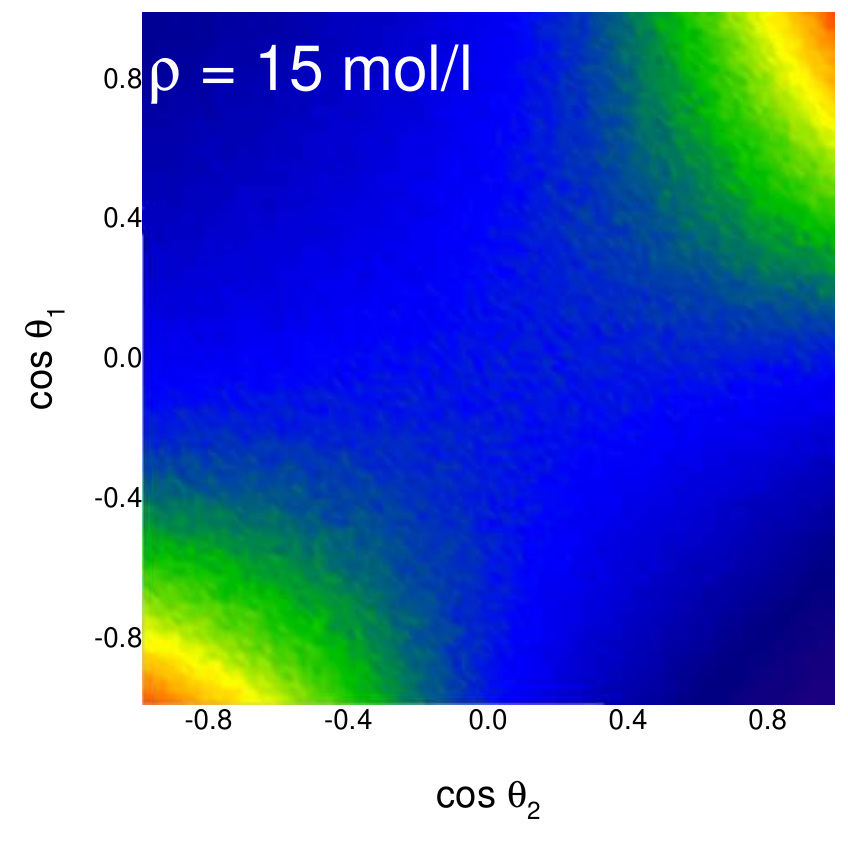}\hfill\includegraphics[width=68mm]{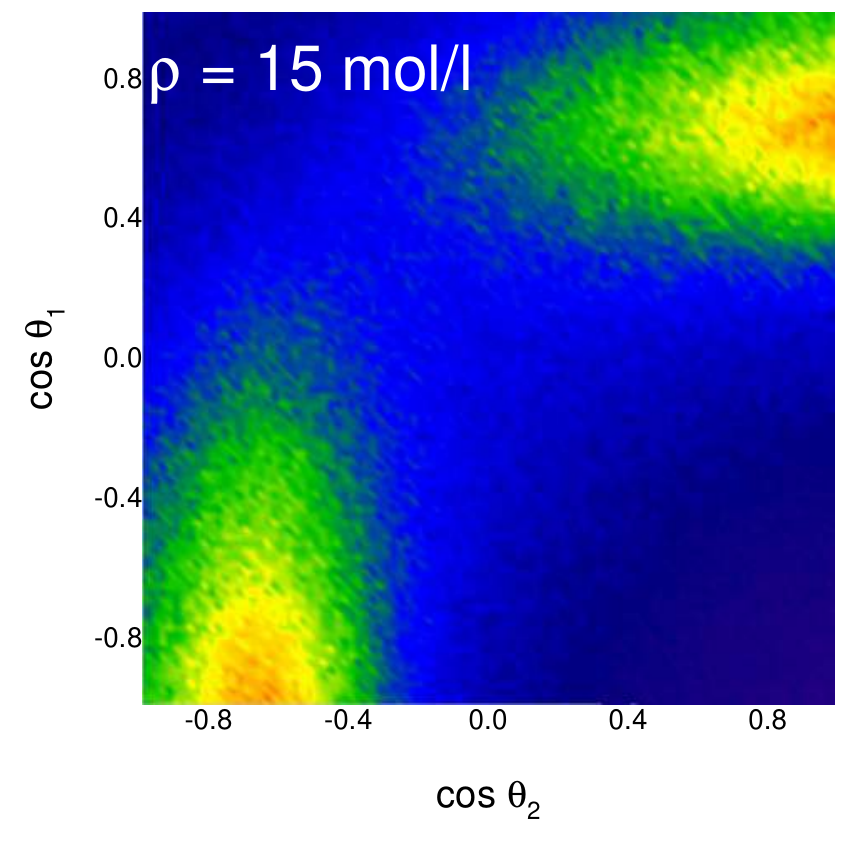}
  {~}
 \end{minipage}

 \begin{minipage}[t]{1.0\textwidth}
	\centering
  \includegraphics[width=68mm]{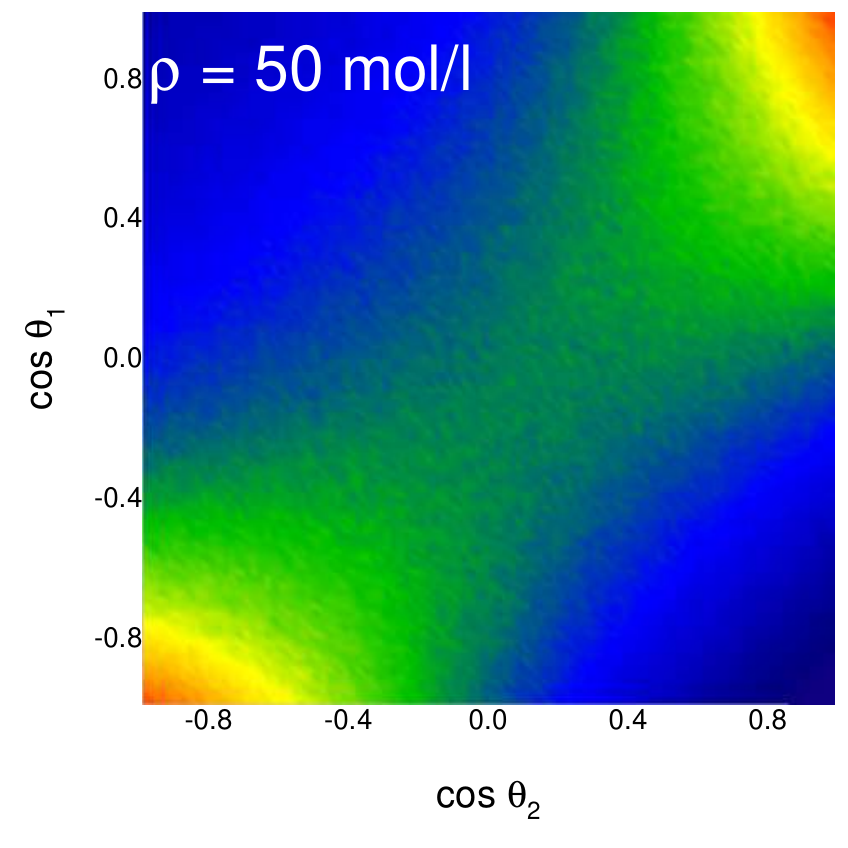}\hfill\includegraphics[width=68mm]{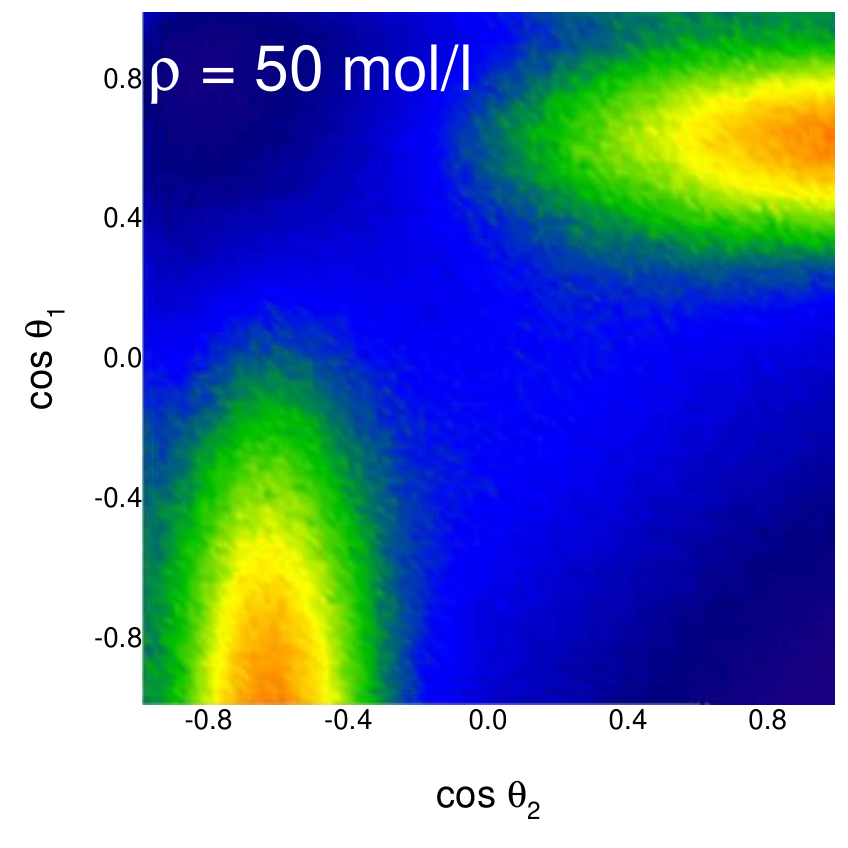}
 \end{minipage}
  \caption{Dipole moment correlation function for the first solvation shell for the 
	 Stockmayer fluid, left column; QDO water, right column at $T^*=1.036T_\text{c}$ at 
	 denoted densities.}
	\label{Figs:MDF}
\end{figure}

\subsection{Polarizable Stockmayer model}
The original Stockmayer model, consisting of the Lennard-Jones (LJ) isotropic part 
and embedded point dipole, provides one of the simplest models for polar liquids.
In particular, we study a dipole polarizable version using the parameters of 
Ref.~\cite{Leeu94} as a baseline. In order to run the simulations in the NAMD 
code \cite{NAMD}, we replaced the point dipole by a pair of $\pm0.52|e|$ point 
charges 0.6~\AA\ apart, giving the permanent dipole moment of 1.5~D. 
A classical Drude model embedded to generate induction in the dipole limit. 
The oscillator consists of a point charge of $-0.5|e|$ harmonically tethered 
to the midpoint between the charges forming the fixed dipole; the latter bears 
a neutralizing charge of $+0.5|e|$. The generally accepted value for the water 
polarizability, $\alpha=1.444\,\text{\AA}^3$, defines the classical Drude 
spring constant, $k=240\,\text{kJ mol}^{-1}\,\text{\AA}^{-2}$. Apart from the 
Lennard-Jones terms between the Drude tether sites, with parameters 
$\epsilon=3.1785$~kJ/mol and $\sigma=2.955$~\AA, a regularising LJ repulsion 
is added between the oscillators with parameters $\epsilon=0.36854$~kJ/mol, 
$\sigma=2.955$~\AA.

\subsection{Association topology}
In order to study the short-range orientational correlations of water and the 
Stockmayer fluid, we calculated the probability distrinution function for two 
angles $\theta_i$ $i\in\{1,2\}$ between molecular dipole moments and the vector 
connecting their COMs for the molecules in the first solvation shell (as defined 
by the 1st minima in $g_\text{OO}(r)$ radial distribution functions). The 
results presented in Fig.~\ref{Figs:MDF} show that while water prefers 
tetrahedral coordination, the Stockmayer fluid adopts a linear association 
topology.

\end{spacing}
\bibliography{Super_PRL}